\documentclass[reqno,11pt]{amsart}
\usepackage{amsmath, latexsym, amsfonts, amssymb, amsthm, amscd}
\usepackage{multirow,stmaryrd,color}
\usepackage{graphics,epsf,psfrag}
\DeclareMathOperator{\Var}{\mathrm{var}}

\numberwithin{equation}{section}

\newtheorem{theorem}{Theorem}[section]
\newtheorem{lemma}[theorem]{Lemma}
\newtheorem{proposition}[theorem]{Proposition}

\newtheorem{rem}[theorem]{Remark}

\newcommand{\ind}{\mathbf{1}}

\renewcommand{\tilde}{\widetilde}

\newcommand{\cF}{{\ensuremath{\mathcal F}} }

\newcommand{\cN}{{\ensuremath{\mathcal N}} }

\newcommand{\bP}{{\ensuremath{\mathbf P}} }
\newcommand{\bE}{{\ensuremath{\mathbf E}} }

\DeclareMathSymbol{\leqslant}{\mathalpha}{AMSa}{"36} 
\DeclareMathSymbol{\geqslant}{\mathalpha}{AMSa}{"3E} 
\DeclareMathSymbol{\eset}{\mathalpha}{AMSb}{"3F}     
\newcommand{\dd}{\,\text{\rm d}}             
\newcommand{\sumtwo}[2]{\sum_{\substack{#1 \\ #2}}} 

\newcommand{\bbE}{{\ensuremath{\mathbb E}} }

\newcommand{\bbN}{{\ensuremath{\mathbb N}} }

\newcommand{\bbP}{{\ensuremath{\mathbb P}} }

\newcommand{\bbR}{{\ensuremath{\mathbb R}} }

\newcommand{\bbZ}{{\ensuremath{\mathbb Z}} }

\newcommand{\ga}{\alpha}
\newcommand{\gb}{\beta}
\newcommand{\gd}{\delta}
\newcommand{\gep}{\varepsilon}       

\newcommand{\gr}{\rho}

\newcommand{\gG}{\Gamma}

\newcommand{\go}{\omega}

\newcommand{\gl}{\lambda}
\newcommand{\gL}{\Lambda}
\newcommand{\gs}{\sigma}

\makeatletter
\def\captionfont@{\footnotesize}
\def\captionheadfont@{\scshape}

\long\def\@makecaption#1#2{%
  \vspace{2mm}
  \setbox\@tempboxa\vbox{\color@setgroup
    \advance\hsize-6pc\noindent
    \captionfont@\captionheadfont@#1\@xp\@ifnotempty\@xp
        {\@cdr#2\@nil}{.\captionfont@\upshape\enspace#2}%
    \unskip\kern-6pc\par
    \global\setbox\@ne\lastbox\color@endgroup}%
  \ifhbox\@ne 
    \setbox\@ne\hbox{\unhbox\@ne\unskip\unskip\unpenalty\unkern}%
  \fi
  \ifdim\wd\@tempboxa=\z@ 
    \setbox\@ne\hbox to\columnwidth{\hss\kern-6pc\box\@ne\hss}%
  \else 
    \setbox\@ne\vbox{\unvbox\@tempboxa\parskip\z@skip
        \noindent\unhbox\@ne\advance\hsize-6pc\par}%
\fi
  \ifnum\@tempcnta<64 
    \addvspace\abovecaptionskip
    \moveright 3pc\box\@ne
  \else 
    \moveright 3pc\box\@ne
    \nobreak
    \vskip\belowcaptionskip
  \fi
\relax
}
\makeatother
\def\writefig#1 #2 #3 {\rlap{\kern #1 truecm
\raise #2 truecm \hbox{#3}}}

\newcommand{\tf}{\textsc{f}}

\begin{document}

\title[Disorder and wetting transition for pinned harmonic crystal]{Disorder and wetting transition:  the pinned harmonic crystal in dimension three or larger}

\author{Giambattista Giacomin}
\address{
  Universit\'e Paris Diderot, Sorbonne Paris Cit\'e,  Laboratoire de Probabilit{\'e}s et Mod\`eles Al\'eatoires, UMR 7599,
            F- 75205 Paris, France. 
}

\author{Hubert Lacoin}
\address{
  IMPA, Institudo de Matem\'atica Pura e Aplicada, Estrada Dona Castorina 110
Rio de Janeiro, CEP-22460-320, Brasil. 
}

\begin{abstract}
We consider the Lattice Gaussian free field in $d+1$ dimensions, $d=3$ or larger, on a  large box (linear size $N$) with boundary 
conditions zero.  On this field two potentials are acting: one, that models the presence of a wall, 
penalizes the field when it enters the lower half space and  one, the {\sl pinning potential}, that rewards  visits to  the proximity of the wall. The wall can be soft, i.e. the field
has a finite penalty to enter the lower half plane, or hard when the penalty is infinite. In general the pinning potential is disordered 
and it gives on average a  reward $h\in \bbR$ (a negative reward  is a penalty): the energetic contribution when the  field at site $x$ visits the pinning region is 
$\gb \go_x+h$, $\{\go_x\}_{x \in \bbZ^d}$ are IID centered and exponentially integrable random variables of unit variance and $\gb\ge 0$.
 In \cite{cf:BDZwetting} it is shown that, when $\gb=0$ (that is, in the non disordered model), a delocalization-localization  transition happens at 
 $h=0$, in particular the free energy of the system is zero for $h \le 0$ and positive for $h>0$. 
 We show that, for $\gb\neq 0$, the transition happens at $h=h_c(\gb):=- \log \bbE \exp(\gb \go_x)$ and we
 find the precise asymptotic behavior of the logarithm of the free energy density of the system when $h \searrow h_c(\gb)$.
 In particular, we show that the transition is of infinite order in the sense that the free energy is smaller than any power of $h-h_c(\gb)$ in the 
 neighborhood of the critical point
 and that disorder does not modify at all the 
 nature of the transition. We also provide  results on the behavior of the paths of the random field in the limit $N \to \infty$.
   \\[10pt]
  \\[10pt]
\end{abstract}

\maketitle


\section{Introduction}

The object of this paper is  the nature of the wetting transition for a $d$-dimensional harmonic crystal interacting with a substrate and 
the effect of disorder on  this transition.

The harmonic crystal, or lattice Gaussian free field (LGFF), is the basic model for surfaces with Hamiltonian  given by the sum of the square of the gradients of the field.
Its Gaussian nature makes it, in most of the cases, easier to analyze than other surface fields with gradient potential
and   conclusions drawn for LGFF are expected to remain valid for a larger class of field.

This is  the case for the study of the wetting transition which involves a competition between a repelling potential (possibly infinite) 
acting on the lower half-space and an attracting one located on a band of finite width above this half-space.
What one finds in the literature about this specific problem -- the literature on LGFF is very vast since it naturally emerges  in a variety of contexts see \cite{cf:Vel,cf:ofer} and references therein -- can be resumed as follows:
\smallskip

\begin{itemize}
 \item In the absence of attracting potential, a wall constraint in the lower half-plane induces a phenomenon of repulsion of entropic origin  in dimension $d=2$ and $d\ge 3$.
The surface lies at a distance from the wall which is of order $\log N$ in dimension two and $\sqrt{\log N}$ in dimension three or larger when $N$ is the size of the system \cite{cf:BDZ,cf:D,cf:DG,cf:LM} .
 \item An arbitrary small (positive) {\sl pinning} potential in the intermediate band is sufficient to overcome this entropic repulsion when $d\ge 3$ \cite{cf:BDZwetting}
 whereas when $d\ge 2$, the repulsion prevails even in the presence of a small positive potential \cite{cf:CV}.  So when $d\ge 3$ there is a transition when the potential switches from repulsive to attractive, while the transition happens at some positive value of the pinning potential. 
\end{itemize}
\smallskip

In the present work we analyze the phase transition for $d\ge 3$, with a twofold objective:
\begin{itemize}
 \item We study the free-energy behavior at the vicinity of the critical point and show that the transition is  of infinite order.
 \item We investigate the effect of disorder on this phase transition 
 and show that quenched and annealed critical points coincide.
Moreover we show that  the critical behavior is not modified by the disorder.  
\end{itemize}
We also prove that in the localized phase, the distribution of the field in the middle of large box converges when the boundary is sent  to infinity to a 
translation covariant limiting distribution.

\smallskip 

These results offer a sharp contrast with those obtained in the absence of half-space repulsion \cite{cf:GL, cf:FF2} 
(see also \cite{cf:CM1} for a first contribution to the  subject). In that case, the transition, which is of first order when $d\ge 3$  and
of second order when $d=2$ for the homogeneous case,  becomes smoother in the disordered one (order two and infinity respectively).These difference can be interpreted in the light of Harris criterion concerning disorder relevance \cite{cf:Hcrit}: for the wetting transition here, 
the homogeneous model has a smooth transition (the specific heat exponent is negative), and for this reason, disorder should be irrelevant, i.e. it should not change the critical behavior, at least for small 
perturbation. For the pinning transition studied in \cite{cf:GL, cf:FF2}  the  specific heat exponent is positive,  
so the Harris criterion predicts disorder relevance.

\smallskip

Note that the model is also defined in dimension one: in that case the harmonic crystal is simply a random walk with IID Gaussian increments.
In that case the behavior of the model is quite different and very similar to the random walk pinning model (the case where no wall is present) 
for which an extended literature exists  (see \cite{cf:sohier} for a treatment of the non disordered case and \cite{cf:GB,cf:G}
and references therein for one dimensional disordered pinning models).

\section{Model and results}

\subsection{Wetting models, with and without disorder}
Given $\Lambda$ be a finite subset of $\bbZ^d$ ($\Lambda$ is always going to be an hypercube and $d=3,4, \ldots$), we
 let $\partial \gL$ denote the internal boundary of $\Lambda$ and 
$\mathring{\gL}$ the set of interior points of $\gL$, that is  (with $\sim$ standing for nearest neighbor)
\begin{equation}\label{boundary}
\partial \gL\, =\,\{x \in \gL : \,  \textrm{there exists } y\notin \gL \textrm{ such that } x\sim y  \} \ \ \  \text{ and } \ \ \
\mathring{\gL}\,:=\, \gL \setminus \partial \gL\, .
\end{equation}
$\bP_{\gL}^{\hat \phi}$ is the law of the LGFF on $\Lambda$ 
(denoted by $\phi=\{\phi_x\}_{x\in \bbZ  ^d}$) with boundary conditions $\hat \phi\in \bbR^{\bbZ^d}$ on $\bbZ^d \setminus\mathring{\gL}$.
Explicitly
$ \phi_x= \hat \phi_x$
for $x \notin \mathring{\gL}$ 
and consider $\bP_\gL^{\hat \phi}$ as a probability on $\bbR^{\mathring{\gL}}$ whose density is given by
\begin{equation}
\label{density}
\bP_\gL^{\hat \phi}(\dd \phi) \, \propto \, 
  \exp\left(-\frac 1 2 \sumtwo{(x,y)\in (\gL)^2 \setminus (\partial \gL)^2 }{x\sim y}\frac{ (\phi_x-\phi_y)^2 }{2} \right)
\prod_{x\in \mathring{\gL}} \dd \phi_x \, ,
\end{equation} 
where $\prod_{x\in \mathring{\gL}} \dd \phi_x$ denotes the Lebesgue measure on $\bbR^{\mathring{\gL}}$. 
For the particular case $\hat \phi\equiv u$ we write $\bP^u_\gL$.
In most of the cases 
\begin{equation}
\Lambda\, =\, \Lambda_N\, :=\, \{0,\dots,N\}^d\, ,
\end{equation}
for some (usually large) $N\in \bbN$, so
 $\mathring{\gL}_N:= \{1,\dots, N-1\}^d$. We also introduce the notation $\tilde \gL_N:= \{1,\dots, N\}^d$.

\medskip

\begin{rem}
\label{rem:lgff}
Of course $\bP_\gL^{\hat \phi}$ is the finite volume LGFF. Much has been written about this field:
we stress here that for $d\ge 3$ the $N \to \infty$ limit, with respect to the product topology, of $\bP_{\gL_N}^{u}$ exists and it can be characterized as the Gaussian field with constant expectation $u$ and covariance of $\phi_x$ and $\phi_y$ 
equal to  the expected time spent in $y$ by a simple symmetric random walk issued from $x$ (for more on this very well known issue we refer to \cite[Sec.~2.9]{cf:GL} and references therein). In particular, the variance of $\phi_x$ in the infinite volume limit
does not depend on $x$ and we denote it by $\gs_d^2$. Moreover the random walk representation holds also in finite volume  -- the walk is killed al the boundary -- and this directly implies that the variance of $\phi_x$ grows as the region considered grows in the  sense of set inclusion. 
\end{rem}

\medskip

Given $\go=\{\go_x\}_{x \in \bbZ^d}$ a family of  IID square integrable centered random variables (of law $\bbP$)
with unit variance, we set, for all $\gb\in \bbR$
\begin{equation}
\label{eq:defgl}
 \gl(\gb)\, :=\, \log \bbE[e^{\gb \go_x}] \, .
\end{equation}  
We call $I_{\bbP}$  the interval where $\gl(\gb)$ is finite and assume that it contains a neighborhood of the origin.
The two families of random variables, $\go$ with law $\bbP$ and the LGFF $\phi$ with law $\bP_\gL^{\hat \phi}$,
are realized on a common probability space and they are independent. 

\medskip

For $x\in \bbZ^d$  set  $\delta_x:= \ind_{[0,1]}(\phi(x))$ and $\rho_x:= \ind_{(-\infty,0)}(\phi(x))$.
For $\gb\in I_{\bbP}$, $h\in \bbR$ and $K\in \bbR\cup \{+\infty\}$ (but in the main results $K \in (0, \infty]$), we define  a modified measure $\bP_{N, h,K}^{  \gb,\go , \hat \phi}$ via
\begin{equation}
\label{eq:modmeas}
\frac{\dd \bP^{\gb,\go,\hat \phi}_{N,h,K}}{\dd \bP^{\hat \phi}_N}(\phi)\, =\, \frac{1}{Z^{\gb,\go,\hat \phi}_{N,h,K}}\exp\left( \sum_{x\in  \tilde \gL_N}
\big( (\gb \go_x-\gl(\gb)+h)\delta_x -K \gr_x\big)\right)\, .
\end{equation}
where
\begin{equation}
\label{eq:modZ}
Z^{\gb,\go,\hat \phi}_{N,h,K}\,:=\,\bE^{\hat \phi}_N\left[ \exp\left( \sum_{x\in  \tilde \gL_N} \big(
 (\gb \go_x-\gl(\gb)+h)\delta_x -K \gr_x\big)
 \right)\right]\,.
\end{equation}
In the homogeneous case, $\gb=0$, we just drop from the superscripts  $\gb$ and  $\go$.

\subsection{Main results}
We introduce the  free energy (density) for every $K \in (-\infty, \infty]$, every $\gb\ge 0$ such that $\gl(\gb)< \infty$ and 
every $h \in \bbR$ as 
\begin{equation}
\label{eq:fe-qa}
\tf_K(\gb, h)\, =\, \lim_{N \to \infty} \frac 1{N^d} \bbE \log Z^{\gb,\go,0}_{N,h,K}\, .
\end{equation}
Theorem~\ref{th:fe} ensures that this limit exists, also as an almost sure limit  if we drop the expectation with respect to the disorder. We  note that, from the free energy viewpoint there is no point in paying attention to summing over $\tilde \gL_N$
in the energy term $\left(\sum_{x\in  \tilde \gL_N} \ldots\right)$ defining the partition  function: $\gL_N$ or $\mathring{\gL}_N$ 
give the same free energy. Even more, the measure $\bP^{\gb, \go,\hat \phi}_{N,h}$ does not see these energy changes at the boundary. Where the choice of $\tilde \gL_N$ enters the game in a non negligible way in relation to the super-additive property: at this stage this important issue is just technical (see Appendix~\ref{sec:fe}).

\smallskip

It is very well known that for pinning models the observation  that for every $K\in [0, \infty]$ (and every $\gb\in I_{\bbP}$ and every $h$)
\begin{equation}
\label{eq:fege0}
\tf_K(\gb, h) \, \ge \, 0\,.
\end{equation}
This follows simply from the fact that $-\log\bP_N^0 (\phi_x >1$ for every  $x\in  \tilde \gL_N)=o(N^d)$ \cite{cf:LM}. 
The bound \eqref{eq:fege0} combined with the fact that the convex function $\tf_K(\gb, \cdot)$ is non decreasing, tells us that there exists
$h_{c,K}(\gb)$ (at this stage we drop the dependence on $K$ for conciseness), a priori in $[-\infty, \infty]$,  such that $\tf_K(\gb, h)>0$ if and only if $h>h_c(\gb)$. Elementary arguments
directly yield that $h_c(\gb)\in [h_c(0), h_c(0)+ \gl(\gb)]$ and that $h_c(0)\in [0, \infty)$: $h_c(\gb)\ge h_c(0)$ is just a consequence of the {\sl annealed 
bound} (Jensen inequality) 
\begin{equation}\label{eq:annehilde}
\tf(\gb, h)\, =\, \lim_{N\to \infty} \frac{1}{N}\bbE\left[\log  Z^{\gb,\go,h}_{N,h,K}\right]\le \lim_{N\to \infty} 
\frac{1}{N}\log  \bbE\left[ Z^{\gb,\go,h}_{N,h,K}\right] =  \tf(\gb, 0).
\end{equation}
The bound $h_c(\gb) \le h_c(0)+ \gl(\gb)$ follows by convexity of $\tf(\cdot, h)$, using  $\partial_\gb \tf(0, h)=0$ (see e.g. \cite[p.~23]{cf:GB}). 
Finally $h_c(0)\ge 0$ is just direct consequence of $Z^{0}_{N, h, K}\le 1$ if $h\ge 0$ and 
$h_c(0)\le -\log C:=-\log \bP( \cN \in [0, 2d])$ follows from 
$\bP_N^0 (\phi_x \in [0,1]$ for every  $x\in  \tilde \gL_N) \ge C^{-N^d}$, which one derives by an easy nearest neighbors conditioning argument.

\medskip
\begin{theorem}
\label{th:mainfe}
For every $K\in (0, \infty]$, every $\gb\in I_{\bbP}$ we have 
\begin{equation}
h_c(\gb)\, =\, 0\, .
\end{equation}
Furthermore if $\gb$ belongs to the interior of $I_{\bbP}$ we have that for $h \searrow 0$
\begin{equation}
\label{eq:mainfe}
 \tf_K(\gb, h)\, =\, \exp \left( 
 \left(-\frac {\gs_d^2}2 + o(1) \right)
  \left( \log \frac 1h \right)^2\right)\, .
\end{equation}
\end{theorem}

\medskip

The proof of Theorem~\ref{th:mainfe} is in Section~\ref{sec:beta=0} (Proposition~\ref{th:pK}: case $\gb=0$ and upper bound estimate for 
$\gb>0$) and in Section~\ref{sec:beta>0} (Proposition~\ref{th:lb-disorder}: lower bound estimate when $\gb>0$).

\smallskip

It is worth pointing out that also in the case $K=0$, treated in \cite{cf:GL}, we have that $h_c(\gb)=0$, but 
\eqref{eq:mainfe} does not hold! In fact in \cite{cf:GL} it is shown that the critical behavior in that case 
has a power law behavior. More importantly, in \cite{cf:GL} it is shown that in the $K=0$ case disorder is relevant, i.e.
it changes the critical behavior (for any $\gb>0$) -- while \eqref{eq:mainfe} shows disorder irrelevance (more on this just below).
A less important remark is that 
the case $K<0$ can be mapped to the case $K>0$: the upper half space is in this case penalized and 
it is an easy matter to work out the correspondences.

Moreover Theorem~\ref{th:mainfe} directly generalizes to the case in which $\gd_x$ is defined by 
$b \ind_{[0,a]}(\phi_x)$, with $a$ and $b>0$: \eqref{eq:mainfe} should simply be replaced by
\begin{equation}
\label{eq:mainfe-ab}
 \tf_K(\gb, h)\, =\, \exp \left( 
 \left(-\frac {\gs_d^2}{2a^2} + o(1) \right)
  \left( \log \frac 1h \right)^2\right)\, .
\end{equation}

\smallskip

The novel content of 
Theorem~\ref{th:mainfe}  is twofold
\smallskip

\begin{enumerate}
\item It improves substantially what was known in the literature, and notably the results in \cite{cf:BDZwetting}, where the case
$\gb=0$ has been considered for a pinning potential of the form $b \ind_{[0,a]}(\cdot)$, precisely the one addressed by
\eqref{eq:mainfe-ab}. In our set-up the potential is rather $h\, b \ind_{[0,a]}(\cdot)$, so we can set $b=1$ and $h(>0)$ plays the role of $b$. 
The result in \cite{cf:BDZwetting} can be restated as $h_c(0)=0$ with a lower bound on the free energy that has not 
been made explicit by the authors. However, a close look at their computation gives the lower bound
\begin{equation}
\label{eq:exfromBDZ}
\tf(0,h)\, \ge\, \exp \left(-c_{\alpha} h^{-\alpha} \right)\, ,
\end{equation}
where $\alpha>2$ is a constant which depends on $c_1$ in \cite[Proposition 2]{cf:BDZwetting}.
Hence \eqref{eq:mainfe} improves considerably this bound and provides a matching lower bound. 
In \cite{cf:BDZwetting}  also a singular limit of the model has been considered: we pick up this model at the end of
Section~\ref{sec:outline}.  
\item Our result also covers the disordered case and shows a strong form of disorder irrelevance, in agreement with the Harris criterion, 
see the review of the literature on this issue in \cite[Sec.~5]{cf:G}. In the Harris criterion perspective, Theorem~\ref{th:mainfe} 
finds a parallel in the results on 
 loop exponent one renewal pinning models \cite{cf:AZ}.
\end{enumerate}

\medskip

It is rather straightforward to extract  from the convexity of the free energy that 
the system has a positive density of contacts if and only if $h>h_c(\gb)$ and therefore $h_c(\gb)$ is the critical point for a localization transition.
One can also get a Large Deviation type estimate on the number of contacts in a large volume in the localized regime, like it is done in   \cite{cf:BDZwetting}
for the  $\gb=0$ case. And it is precisely in \cite{cf:BDZwetting} that obtaining pointwise bounds on the field is cited as an open problem.
Here we present a result in this direction. For this we choose to work, only for the next statement (and its proof, see Sec.~\ref{sec:path}),
with $\gL_N:=\{-N, \ldots , N\}^d$ (and $\tilde \gL_N:=\{-N+1, \ldots , N\}^d$ in \eqref{eq:modmeas}). The measures we consider 
are all viewed either as elements of the set of probability measures on $[0, \infty)^{\bbZ^d}$ or $[0, \infty)^{\bbZ^d}$,
both equipped with the product topology and $[0, \infty]$ is equipped with the usual compactified topology. We use $\Theta_x$ for the translation operator, both for $\phi$, that is $(\Theta_x \phi)_y= \phi_{x+y}$, and for $\go$. 
 
\medskip

\begin{theorem}
\label{th:path}
Let us choose $\gb\in I_{\bbP}$. 
\begin{enumerate}
\item If $h>0$  
 the  sequence of {\sl quenched averaged} probabilities  $\{ \bbE \bP^{\go, \gb, 0}_{N, h, \infty}\}_{N=1,2, \ldots}$, probabilities  on $[0, \infty)^{\bbZ^d}$, 
converges to a translation invariant limit. Moreover for $\bbP$-almost every $\go$ the sequence 
$\{\bP^{\go, \gb, 0}_{N, h, \infty}\}_{N=1,2, \ldots}$, probabilities  on $[0, \infty)^{\bbZ^d}$, converges to a translation covariant limit
$\bP^{\go, \gb, 0}_{\infty, h, \infty}$, that is $\bE^{\go, \gb, 0}_{\infty, h, \infty}[f(\Theta _x \phi)]=\bE^{\Theta_x\go, \gb, 0}_{\infty, h, \infty}[f(\phi)]$ for every bounded local $f$. 
\item If $h \le 0$   the sequence of quenched averaged and quenched  measures,  both probabilities  on $[0, \infty]^{\bbZ^d}$,
converge to the probability concentrated on the singleton  $\{\infty\}^{\bbZ^d}$.
\end{enumerate}
\end{theorem}

\begin{rem}
 Let us observe here that the proof of Theorem \ref{th:path} does not rely much on the assumption $d\ge 3$.
 When $d=2$, a non-trivial covariant limit exists when $h>h_c(\gb)$ (as a consequence of \cite{cf:CV} and of the annealed bound \eqref{eq:annehilde} 
 $h_c(\gb)\ge h_c(0)>0$ for all $\gb$)
 while for  $h<h_c(\gb)$, the limit is concentrated on  $\{\infty\}^{\bbZ^2}$ (the proof is identical). We cannot conclude in the case $h=h_c(\gb)$
since it is not known whether the localization transition is of first order or higher.
In dimension one, results of the same type have been known for a long time (see \cite[Section 7.3]{cf:G})
\end{rem}

The proof of Theorem~\ref{th:path} is in Section~\ref{sec:path}.

\subsection{Outline of the proof of Theorem~\ref{th:mainfe}}
\label{sec:outline}
We will first treat, in Section~\ref{sec:beta=0},  the homogeneous, or pure, model (i.e., $\gb=0$).
This both for presenting the easier case first and because $\tf_K(\gb, h)\le \tf_K(0, h)$ so, in Section~\ref{sec:beta>0} dedicated to $\gb>0$, we just need to provide a lower bound 
 on  $\tf_K(\gb, h)$. 
 
 The key point behind Theorem~\ref{th:mainfe} is that a one site strategy turns out to be sufficient. Roughly the idea is the following: 
 We can imagine that close to criticality, the field has very few pinned sites (note that this would not be the case if the 
 transition were of first order, but as we are still at the stage of making guesses and it does not seem unreasonable to think that the transition is smooth). 
 Therefore, also helped by the (entropic) repulsion 
 effect of the the  (soft, $K< \infty$, or hard, $K=\infty$) wall and by the massless character of the field, the field is expected to be  at a typical height $u\gg 1$, hence quite far from the levels that contribute to the energy. So the energy contributions are due to rare spikes downwards. The various sharp results on 
entropic repulsion  for LGFF in  $d\ge 3$, notably \cite{cf:BDZ,cf:D,cf:DG}, support this intuition and also the fact that 
large excursions of the LGFF in  $d\ge 3$ are essentially just isolated spikes (see \cite{cf:CCH} for a convergence result of these spikes to a Poisson process): for example, the expectation of $\max_{x \in \gL_N}
\phi_x$, where $\phi$ the infinite volume centered   
LGFF in  $d\ge 3$, is, to leading order as $N \to \infty$, the same as if $\phi$ were a collection of IID $\cN(0, \gs_d^2)$ random variables (recall that $\gs_d^2$ is the variance of the one dimensional marginal of the infinite volume LGFF). 
So let us imagine that the field is repelled for $h$ small to a height $u$ very large and  that we can look at the contribution
of each variable like if they were independent. So we are reduced to the computation of the contribution to the partition function of one site in this idealized setup:
\begin{multline}
\label{eq:onesite1}
\bP(\phi_x >1)+ e^h \bP(\phi_x \in [0,1])+ e^{-K} \bP(\phi_x <0)\, = \\
1+ (e^h-1) \bP(\phi_x \le 1)+ (e^{-K}-e^h) \bP(\phi_x <0)\, .
\end{multline} 
Now, we use $\phi_x\sim \cN(u, \gs_d^2)$,
so $\phi_x= \gs_d \cN +u$ ($\cN$ is a standard Gaussian variable, or $\cN\sim \cN(0,1)$) and 
 the standard asymptotic estimate
\begin{equation}
\label{eq:Nasympt}
\bP(\cN>t) \stackrel{t\to \infty}\sim \frac1{t \sqrt{2\pi}}\exp\left(-\frac{t^2}2\right)\, .
\end{equation}
Since $h$ is small and $u$ is large we can approximate  \eqref{eq:onesite1} 
by 
\begin{multline}
\label{eq:onesite2}
1+ \frac{h \gs_d}{u \sqrt{2\pi}}\exp\left(-\frac {(u-1)^2}{2 \gs_d^2}\right) - \left(1-e^{-K}\right) 
\frac{ \gs_d}{u \sqrt{2\pi}}\exp\left(-\frac {u^2}{2 \gs_d^2}\right)\, =\\
1+ \frac{ \gs_d}{u \sqrt{2\pi}}\exp\left(-\frac {u^2}{2 \gs_d^2}\right) \left[
h \exp\left(\frac {u}{ \gs_d^2}-\frac {1}{2 \gs_d^2}\right)
- \left(1-e^{-K}\right)
\right]\,.
\end{multline}
Up to now we have not said anything about the value if $u$, but this computation says that a positive contribution to the free energy 
requires  
 the term
in the square brackets to be positive. And  for this one needs to choose 
$u= \gs_d^2 \log (1/h)+c$ for some positive constant $c$ and choosing a much larger $u$ would strongly 
penalize the gain, because of the prefactor $\exp(-{u^2}/(2 \gs_d^2))$. 
Note the marginal role played in this computation
by $K$, as long as $K>0$. This computation is suggesting that the one site contribution is for $h \searrow 0$
\begin{equation}
\label{eq:onesite3}
1+ \exp\left(-\frac {(\gs_d^2+o(1)) \left( \log\frac 1h\right)^2}{2 }\right)\, ,
\end{equation}
and therefore 
\begin{equation}\begin{split}
\tf_K(0,h) &\, =\, \log \left( 1+ \exp\left(-\frac {(\gs_d^2+o(1))  \left( \log\frac 1h\right)^2}{2 }\right)\right)\\
            &\, =
\, \exp\left(-\frac {(\gs_d^2+o(1))  \left( \log\frac 1h\right)^2}{2 }\right)\, .
\end{split}\end{equation}
This is the main claim of Theorem~\ref{th:mainfe} and in order to convert such an argument into a proof we
will proceed separately for upper and lower bound. The upper bound is achieved by reducing the estimate to a model on a (very) spaced sub-lattice (via an application
of the H\"older inequality): the Markov property of the LGFF at this point can be used to provide enough independence to obtain the bound we are after. 
For the lower bound we exploit the fact (in Appendix~\ref{sec:fe}) that the free energy can be computed by choosing boundary conditions that are sampled from the infinite volume
LGFF with an arbitrary average height $u$: 
in fact, with such boundary conditions the logarithm of the partition function 
forms a super-additive sequence, hence we can perform estimates for finite $N$ to estimate from below the $N=\infty $ case,
see \eqref{eq:superadd}. 
We then proceed 
 by using Jensen's inequality in a way that a priori may seem very rough (we just compute the expectation of the energy term!), 
 but this turns ou to be sufficient, thanks to the first step and the wise choice of the boundary mean $u$.
 
For what concerns the disordered case, the desired lower bound on  $\tf_K(\gb,h)$
is achieved  once again by exploiting super-additivity -- we will  work with a finite volume size that diverges as $h$ tends to zero -- 
and by choosing the boundary values at average height $u$. We choose $u=\gs_d^2 \log (1/h)+c$. It was argued after  \eqref{eq:onesite2}
that this should be the value of $u$ that maximizes the energy gain in the pure case and we 
choose to use the same value for the disordered case because we are aiming at showing disorder 
irrelevance. 
The volume size is chosen so that it is improbable to observe two or more contacts and then the estimate 
is performed on the partition function limited to the trajectories of the field that have at most one contact (and we send 
also $K$ to infinity, since, by monotonicity in $K$ of the partition function, this is the worst case scenario). 
On this reduced free energy we perform a second moment argument.  A look at the proof shows that the variance term 
in this computation plays a very marginal role, reinforcing the idea that disorder is very irrelevant in this model. Even more,
we are able to apply the second moment method without assuming that the second moment of $e^{\gb\go_n}$ is finite!
This is achieved with an accurate cut-off procedure. 
The core of the lower bound argument  on $\tf_K(\gb,h)$ 
also for $\gb\neq 0$ is in any case the one site computation we have just sketched and the argument 
in Section~\ref{sec:beta>0} can be used verbatim (just set $\gb=0$) to obtain another (somewhat more involved) proof  of the lower bound presented in Section~\ref{sec:beta=0} for the non disordered  case. 

\smallskip

We complete this section by pointing out that our arguments  yield partial results about the $\gb=0$ case treated in \cite{cf:BDZwetting} under the name of
 $\gd$-pinning model. This corresponds to the $a\searrow 0$ case of the $b\ind_{[0,a]}(\cdot)$ potential of \eqref{eq:mainfe-ab}, with 
 $h=1$, $b= \exp(c_a+J)$ with $J\in \bbR$ and $c_a$ such that $\lim_{a\searrow 0} a \exp(c_a)=1$. The limit model has a number of nice features, but the model is not critical at any value of $J$:  delocalization arises only for $J \to-\infty$. 
And in fact our results show that, with such parametrization, for $a>0$ the critical value is $J_c=-c_a$, with the corresponding 
critical behavior directly readable from  \eqref{eq:mainfe-ab}. Our arguments of proof suggest that the free energy 
of the $\gd$-pinning model as $J \to -\infty$ is equal to 
\begin{equation}
\exp\left(-\frac 12\left(\gs_d^2+o(1)\right) \exp(-2J)\right)\, .
\end{equation}
The upper bound part of this conjecture can be proven in a straightforward way by the arguments we use, but our lower bound
arguments do not yield the result because they
require a super-additive statement like Proposition~\ref{th:superadd} for the $\gd$-pinning model. On the other hand, in \cite{cf:BDZwetting}
no explicit bound is given, but, like for \eqref{eq:exfromBDZ}, reconsidering their approach we find a lower bound
for the free energy of $\exp(-\exp(-2\ga J))$, for $J<0$ large and $\ga$ a constant larger than $2$.


\section{The homogeneous case}
\label{sec:beta=0}

The main result of this section, Proposition~\ref{th:pK},  implies Theorem~\ref{th:mainfe} for $\gb=0$ and provides the upper bound for the case $\gb>0$.
\medskip 

\begin{proposition}
\label{th:pK}
For every $K\in (0, \infty]$ in the limit $h \searrow 0$ we have
\begin{equation}
 \tf_K(0,h)\, =\, \exp \left( 
 \left(-\frac {\gs_d^2}2 + o(1) \right)
  \left( \log \frac 1h \right)^2\right)\, .
\end{equation}
\end{proposition}

\medskip

\noindent
{\it Proof.} We treat separately the upper and lower bound.

\smallskip

\subsubsection{Upper bound} Since the partition function decreases as $K$ increases,
for the upper bound it suffices to prove the statement for a $K>0$. It is also sufficient to consider $N=n L-1$, with $n, L \in \bbN$, $L$ even (both sufficiently large, say larger than $3$ at this stage, but later on $L$ is chosen fixed but arbitrarily large and $n$ is sent to $\infty$) and the quantity
\begin{equation}
\label{eq:ZnL}
Z_{n,L}\, :=\, \bE_N^0
\exp\left( \sum_{x\in \{L,L+1, \ldots, (n-1)L-1\}^d}
\left( h\, \delta_x -K \gr_x\right)\right)\, .
\end{equation}
Note that the sum in the exponential does not range over the whole box $\tilde \gL_N$, but  there are only $O(n^{d-1}L^d)$ terms missing
and for this reason 
for every fixed $L$ we have 
\begin{equation}
\lim_{n\to \infty} \frac{1}{(Ln)^d} \log Z_{n,L}=\lim_{n\to \infty} \frac{1}{(Ln)^d}  \log Z^0_{nL-1,h,K}= \tf(0,h).
\end{equation}
We now set for $v\in \{0, \ldots, L-1\}^d=:B_L$
\begin{equation}
 \gL_{L,n}^v\, :=\, (v+L \bbZ^d)\cap  \{L,L+1, \ldots, (n-1)L-1\}^d \, ,
\end{equation}
and  the family $\{\gL_{L,n}^v\}_{v\in B_L}$ is a partition of
$ \{L,L+1, \ldots, (n-1)L-1\}^d$ and 
\begin{multline}
Z_{n,L}\, =\, \bE_N^0\left[
\exp\left( \sum_{v\in B_L}\sum_{x\in \gL_{L,n}^v}
\left( h\, \delta_x -K \gr_x\right)\right)\right]\\ 
\le\,
 \prod_{v\in B_L}
\left(\bE_N^0 \left[\exp\left( L^d\sum_{x\in \gL_{L,n}^v}
\left( h\, \delta_x -K \gr_x\right)\right)\right]\right)^{L^{-d}},
\end{multline}
 by H\"older inequality.

Next, we condition on  $\{\phi_y\}_{y\in \gG^v_{N,L}}$ where 
\begin{equation}
\gG^v_{N,L}\,  :=\, \left\{ y \in \gL_N \, : \,  \text{there exists } x \in \gL^v_{N,L} \text{ such that }  \max_{i=1^d}\vert (y-x)_i\vert= (L/2) \right\}\, ,
\end{equation}
 is just a grid that separates the sites
$x\in \gL_{L,n}^v$. By  the Markov property of the LGFF
 we readily see that $\{\phi_x\}_{x\in \gL_{L,n}^v}$ is a family of conditionally independent Gaussian variables. Their (conditional) mean is given by the harmonic extension of  
 $\{\phi_y\}_{y\in \gG^g_{N,L}}$ to the full box and their variance is equal to $c^2_L$ the variance of free field with zero boundary conditions 
 in the center of a box of side length 
 $L$, so $\lim_{L\to \infty} c_L=\sigma_d$. Hence we obtain that almost surely 
\begin{multline}
\label{eq:m34}
\bE_N^0 \left[ \exp\left( L^d\sum_{x\in \gL_{L,n}^v}
\left( h\, \delta_x -K \gr_x\right)\right)\, \Bigg| \, y\in \gG^g_{N,L} \right]\, \le \\ 
 \left(
\sup_{u\in \bbR} \bE \exp\left(L^d h \ind_{[0,1]}\left(c_L\, \cN +u\right)- L^dK \ind_{(-\infty,0)}\left(c_L\,\cN +u\right)\right)
\right) ^{\left \vert \gL_{L,n}^v \right\vert }\, ,
\end{multline} 
which yields the same bound for the unconditional expectation.
With the notation $P(a,b)=\bP(\cN \in (a,b))$ we have
\begin{multline}
\label{eq:3Lsup}
\sup_{u\in \bbR} \bE \exp\left(L^d h \ind_{[0,1]}\left({c_L}\,\cN +u\right)- L^dK \ind_{(-\infty,0)}\left({c_L}\,\cN +u\right)\right)
\,=\\  1+
\sup_{u\in \bbR} \left((\exp(L^dh)-1) P(-u,-u+1/c_L)- (1-\exp (-L^d K))P (-\infty, -u)\right)\\
\le \, 1+ \sup_{u\in \bbR} 
\left(2L^d h 
P(-u,-u+1/c_L)- \frac 12  P (-\infty, -u) \right)
 \,,
\end{multline}
where the last step we have used  $h\le L^{-d}$ and $(1-\exp (-L^d K))\ge 1/2$ (so we assume $L$ larger than a suitable constant dependent on $K$).
We note that the argument of the supremum in
the last line in \eqref{eq:3Lsup} is larger than zero if and only if 
\begin{equation}\label{eq:pisitiv}
\frac{P(-u,-u+1/c_L)}{P (-\infty, -u)}\, > \,  \frac1 {4L^d h}\, .
\end{equation}
We now observe that  the function 
\begin{equation}
\label{eq:PoverP}
(-\infty, \infty) \ni u \mapsto
\frac{P(-u,-u+a)}{P (-\infty, -u)} \stackrel{u\to \infty}\sim \exp\left( au-\frac {a^2}2\right)\, ,
\end{equation}
is smooth and positive. Moreover  it goes to zero as $u \to -\infty$ and to $\infty$ as $u$ goes to $+\infty$
(we cite as a fact that  this function is increasing, but we do not use it in our proof).
Therefore
 for $h$ small the supremum will be achieved for $u$ large: in particular \eqref{eq:pisitiv} implies that we can restrict the supremum to 
 \begin{equation}
 u\ge \frac{1}{2c_L}+c_L \log \left(\frac 1 {5L^d h}\right)=u_0(h,L)
 \end{equation}
 By using this information we could get a sharp estimate on the supremum, but we will content ourselves with a much simpler estimate which is sufficient for our purposes. In fact, neglecting the negative term in \eqref{eq:3Lsup}, we obtain that given $L$
 and $a<c^2_L$/2 for $h$ sufficiently small we have  
\begin{multline}
\sup_{u\ge u_0} 2L^d h P(-u,-u+1/c_L)=2L^d h P(-u_0,-u_0+1/c_L)\\
\le \frac{3 L^d h}{\sqrt{2\pi}} \exp \left( -\frac{u_0^2}{2} \right)\le \exp\left(- a ( \log  (1/h))^2 \right).
\end{multline}

Therefore, going back to \eqref{eq:m34}, we have
\begin{equation}
\label{eq:fm34}
\bE_N^0 \left[\exp\left( L^d\sum_{x\in \gL_{L,n}^v}
\left( h\, \delta_x -K \gr_x\right)\right)\right]\, \le \,
 \left(1+
\exp\left(- a \left( \log  \frac 1h\right)^2 \right)
\right) ^{\left \vert \gL_{L,n}^v \right\vert }\, ,
\end{equation} 
and from \eqref{eq:ZnL}
\begin{multline}
\frac 1{(nL)^d}\log Z_{n,L}\, \le \, \frac{\left\vert \gL_{L,n}^v \right\vert}{(nL)^d} \log \left(1+
\exp\left(- a \left( \log  \frac 1h\right)^2 \right)
\right) \\
\, \le \, \frac 2{L^d} \exp\left(- a \left( \log  \frac 1h\right)^2 \right)\, ,
\end{multline}
again for $h$ sufficiently small. By recalling that
$c_L$ can be chosen arbitrarily close to $\gs_d$ we see that  the proof of the upper bound is complete.

\smallskip

\subsubsection{Lower bound}
Also for the lower bound we work with  $K\in (0, \infty)$, but this time we follow the $K$ dependence of the bound. By  Proposition~\ref{th:superadd}, precisely \eqref{eq:superadd},
and Jensen inequality we  obtain that  for every $u$ and every $N$
\begin{equation}
\begin{split}
\tf_K(h)\, &\ge \,  \frac 1{N^d}  \hat \bE^u \bE_N^{\hat \phi} \left[
\sum_{x\in  \tilde \gL_N}
\left( h \, \delta_x -K \gr_x\right)
\right]\, =\, \bE^u \left[ h \, \delta_x -K \gr_x\right]\, \\
&=\,
 \bE \left[h \ind_{[0,1]}\left(\gs_d\,\cN +u\right)- K \ind_{(-\infty,0)}\left(\gs_d\,\cN +u\right)\right]\,.
\end{split}
\end{equation}
Therefore, with the notation used for the upper bound we have that for every $u$
\begin{equation}
\tf_K(h)\,\ge \, h P(-u, -u+1/\gs_d) -K P(-\infty, -u)\, .
\end{equation}
We set $u= \gs_d \log (1/h)+r$, with $r$  to be determined. Therefore for $h$ sufficiently
small (how small depends here on $r$ since we require $u \ge C$ for some deterministic $C$ to use the asymptotic statement \eqref{eq:Nasympt})
\begin{equation}
\begin{split}
\tf_K(h)\,&\ge \, \frac{1}{u \sqrt{2\pi}}\left(\frac h 2 \exp\left(-\frac 12 \left(u-\frac 1{\gs_d}\right)^2\right) -2K \exp\left(-\frac 12 u^2\right)\right)
\\
&=\,
\frac{2h^{\gs_d r}\exp(-r^2/2)}{(r+ \gs_d \log(1/h)) \sqrt{2\pi}}  \exp\left(-\frac {\gs_d^2}2 \left( \log \frac 1h \right)^2 \right)
\left(\frac{e^{-1/(2\gs_d^2)+r/\sigma_d}}4  -K \right) \, .
\end{split}
\end{equation}
We now set $r=\frac 1{2\gs_d}+ \gs_d\log(4(K+1))$  and we get to the explicit bound
\begin{equation}
\label{eq:expl1}
\tf_K(h)\, \ge \, 
\frac{2h^{\frac 12 +\gs_d^2 \log (4(K+1))}\exp\left(-\frac 12 \left(\frac 1{2\gs_d}+ \gs_d\log(4(K+1))\right)^2\right)}
{\left(\frac 1{2\gs_d}+\gs_d \log(4(K+1))+ \gs_d \log(1/h)\right) \sqrt{2\pi}}  
e^{-\frac {\gs_d^2}2 \left( \log \frac 1h \right)^2 }\, .
\end{equation}
Therefore for every  $b>\gs_d^2/2$ and every $K\in (0, \infty)$ there exists $h_0>0$ such that for every $h\in (0, h_0)$
\begin{equation}
\label{eq:lbwithb}
\tf_K(h)\, \ge \,  \exp\left(-b \left( \log \frac 1h \right)^2 \right)\,.
\end{equation}
This completes the proof of the lower bound, except for the case $K=\infty$.

\medskip 

For the lower bound in the case $K=\infty$ we observe that for $K$ large $r$ becomes large too and 
\eqref{eq:expl1} holds for arbitrary $h_0>0$ as $K \to \infty$ (but in our formulas we want to have $\log (1/h)\ge 0$ so
$h_0=1$). 
Now remark that  if we choose 
$K+1=\exp\left( (\log(1/h))^{1/2}\right)$ the ratio in right-hand side of  \eqref{eq:expl1} is bounded 
below by $\exp(- (\log(1/h))^a)$ for any $a>3/2$ and $h$ sufficiently small. So
 for every $b>\gs_d^2/2$ there exists $h_1>0$  such that for every $h\in (0, h_1)$
\eqref{eq:lbwithb} holds.
The conclusion is then an immediate consequence of Lemma~\ref{th:Klarge}, because $\tf_\infty(0, h) \ge 
\tf_K(0, h) - \exp(-K)$.
This completes the proof of the lower bound and therefore the proof of 
Proposition~\ref{th:pK} .
\qed

\section{The disordered case}
\label{sec:beta>0}

\begin{proposition}
\label{th:lb-disorder}
For every $\gb\in I_{\bbP}$ we have $h_c(\gb)=0$. Moreover if
 $\gb$ is in the interior of $I_{\bbP}$, for every $K\in (0, \infty]$ and every $\gep >0$ there exists $h_0$ such that
\begin{equation}
 \tf_K(\gb,h)\, \ge \, \exp  
 \left( -(1+\gep) \frac{\gs_d^2}2
  \left( \log \frac 1h \right)^2
  \right)\, ,
\end{equation}
for $h\in(0, h_0)$. 
\end{proposition}

In this section we set 
\begin{equation}
\xi_x\, :=\, e^{\gb\go_x-\gl(\gb)}\, ,
\end{equation}
and let $\xi$ denote a variable which has the same distribution as all the $\xi_x$.
Note that $\bbE \xi=1$ and 
that the assumption that   $\gb$ is in the interior of $I_\bbP$ is equivalent to 
\begin{equation}
\label{gammatail}
\text{ `` There exist } C>0 \text{ and } \gamma>1 \text{ such that }
 \bbP\left(\xi \, \ge\, t\right) \, \le\,  Ct^{-\gamma} \text{ for every $t\ge 0$.''}
\end{equation}

\medskip

We also use the notation $\rho_x^+:=\ind_{(-\infty, 1]}(\phi_x)$.
For $\tilde a>1$  we set 
\begin{equation}
\label{eq:parameters}
u\, :=\, \tilde a \, \gs_d^2 |\log h| \  \  \text{ and } \  \ N\, =\, \exp(|\log h|^{3/2})\, .  
\end{equation}
A basic recurrent quantity in the proof is going to be 
\begin{equation}
\label{eq:P(u)}
P(u)\, :=\, \bP^u \left( \phi_0 \le 1\right)\,
\begin{cases}
\stackrel{h \searrow 0}\sim
\frac {\gs_d}{u\sqrt{2\pi} } \exp\left(- \frac{(u-1)^2}{2\gs_d^2}
\right)\, ,
\\
 \ge \,\exp\left(-(1+\gep) \frac{\tilde a^2 \gs_d^2}{2} |\log h|^2\right) \, ,
 \end{cases}
\end{equation}
where we have used \eqref{eq:Nasympt} and the inequality, which  holds for every $\gep>0$ and $h$ sufficiently small,
is directly obtained by inserting the value of $u$.
 Moreover for every $c<1$ we have  $\bP^u ( \phi_0 \in [c,1])\sim P(u)$, for $u \to \infty$. Another 
  relevant estimate is 
\begin{equation}
\label{eq:ratio}
 \frac{ \bP^u \left( \phi_0 \le 1\right)}{ \bP^u \left( \phi_0 <0\right)} \stackrel{h\searrow 0} \sim h^{-\tilde a} \exp(-1/(2\gs^2_d))\, .
\end{equation}

Here and in the remainder of the proof, we avoid insisting on the fact that we should choose $h$ such that $\exp(|\log h|^{3/2})\in \bbN$:
obtaining the estimate along this subsequence yields the claim for $h\searrow 0$ by a direct estimate 
and using that $\tf_K(\gb,\cdot)$ is non decreasing. Alternatively one can carry along the proof $N= \lfloor \exp(|\log h|^{3/2})\rfloor$
and deal with the little nuisances that arise.  

We introduce also the event 
\begin{equation} 
G_u\,:=\, \left\{ \phi \in \bbR^{\bbZ ^d}: \, \phi_x> \frac u2 \text{ for } x \in \partial \gL_N\right\}\, .
\end{equation}
The following statement controls the contribution of the 
 bad boundary configurations: 

\medskip

\begin{lemma}
\label{th:Gc}
For every $d$
there exist $C_d>0$ and $h_0>0$ such that   for every $K\ge 0$, $\gb\in I_\bbP$  and for $h\in [0, h_0)$  we have
\begin{equation}
\bbE \hat \bE^u \left[ \left(\log Z^{\gb,\go,\hat \phi}_{N,h,K} \right) \, \ind_{ G_u^\complement}(\left( \hat \phi \right)\right]\,
\ge \, 
- C_d \left(\gl(\gb) \vee K\right)N^{d-1} P(u)\, .
\end{equation}
\end{lemma}

\medskip 
 
\noindent
{\it Proof.} 
 By Jensen inequality
\begin{equation}
\label{eq:byJ-1}
\begin{split}
\bbE \hat \bE^u &\left[  \left(\log Z^{\gb,\go,\hat \phi}_{N,h,K} \right) \, \ind_{ G_u^\complement}\left( \hat \phi \right)\right]\, \\
&\ge \, (-\gl(\gb)+h) 
\bE^u \left[ \sum_{x\in  \tilde \gL_N} \gd_x  ; \, G_u^\complement\right] -K 
\bE^u \left[ \sum_{x\in  \tilde \gL_N} \rho_x  ; \, G_u^\complement\right]
\\
&\ge \, - \left(\gl(\gb) \vee K\right) \bE^u \left[ \sum_{x\in  \tilde \gL_N} \rho_x^+ ; \, G_u^\complement\right]\, ,
\end{split}
\end{equation}
with the notation $\bE^u[\, \cdot\, ; F]= \bE^u[\, \cdot\,  \ind_{F}(\phi)]$.
By  the union bound and by making  an elementary splitting for a $C>0$ we have
\begin{multline}
\label{eq:UnB1}
 \bE^u \left[ \sum_{x\in  \tilde \gL_N} \rho_x^+ ; \, G_u^\complement\right]\,
 \,\le \, 
 \sumtwo{x\in  \tilde \gL_N}{y \in \partial \gL_N} \bP^u \left( \phi_x \le 1, \, \phi_y \le u/2 \right)
 \\
  \le\, 
 \sumtwo{x\in  \tilde \gL_N, \, y \in \partial \gL_N:}{\vert x-y\vert \le C} \bP^u \left( \phi_x \le 1, \, \phi_y \le u/2 \right)
 +\sumtwo{x\in  \tilde \gL_N, \, y \in \partial \gL_N:}{\vert x-y\vert > C} \bP^u \left( \phi_x \le 1, \, \phi_y \le u/2 \right)
 \\ \le \, 2dC \, N^{d-1} P(u) + \sumtwo{x\in  \tilde \gL_N, \, y \in \partial \gL_N:}{\vert x-y\vert > C} \bP^u \left( \phi_x \le 1, \, \phi_y \le u/2 \right)
 \,.
 \end{multline}
 Now given $\eta>0$ we have 
 \begin{equation}
  \bP^u \left( \phi_x \le 1, \, \phi_y \le u/2 \right)\,\le\,   \bP^u \left( (\phi_x +\eta \phi_y) \le  1+ u\eta/2 \right)\, .
 \end{equation}
Now  $\phi_x +\eta \phi_y$ is a Gaussian variable of mean $(1+\eta) u$.
To compute its variance we observe that the covariance between $\phi_x$ and $\phi_y$ is given by $G(x,y)=\sigma_d^2p(x,y)$  
with $p(x,y)$ the probability that a simple random walk issued from $x$ hits $y$: 
since $p(x,y)$ vanishes when $\vert x-y\vert$ becomes large, we choose $C$ so that $p(x,y)\le 1/8$ when $|x-y|\ge C$.
Hence we have for $\eta\le 1/4$
\begin{equation} 
\Var (\phi_x +\eta \phi_y)\,\le\,  \sigma^2_d (1+\eta^2 +\eta/4)\, \le\,   \sigma^2_d (1+\eta/2)\, .
\end{equation}
Using this information we have  for $u$  sufficiently large 
\begin{multline}
 \bP^u \left( (\phi_x +\eta \phi_y)\le  1+ u\eta/2 \right)\,\le\,  \exp\left(-\frac{(u(1+\eta/2)-1)^2}{\gs^2_d(2+\eta)}\right)\\
 \le P(u) \exp(-c(\eta)u^2)\, \le \,  N^{-d}P(u)\, ,
\end{multline}
where $c(\eta)= \eta/(8\gs^2_d)$ and we have used 
the 
first line in \eqref{eq:P(u)}
 together with the relation between the parameters \eqref{eq:parameters}.

Hence, going back to \eqref{eq:UnB1}, we see that
\begin{equation}
\label{eq:fromUnB1}
 \bE^u \left[ \sum_{x\in  \tilde \gL_N} \rho_x^+ ; \, G_u^\complement\right]\,
 \,\le \, 
 \left(2d(C+1) \, N^{d-1} \right) P(u)
\, .
\end{equation}
By plugging this estimate into \eqref{eq:byJ-1} we complete the proof.
\qed

\medskip

\noindent
{\it Proof of Proposition~\ref{th:lb-disorder}}.
We aim at producing a lower bound on $Z^{\gb,\go,\hat \phi}_{N,h,K}$ 
for good boundary values $\hat \phi$ (Lemma~\ref{th:Gc} is going to take care of the bad ones). This will be achieved 
by a second moment approach: we give first the proof assuming that the second moment of $\xi$ is finite, that is  
$\gl (2\gb)<\infty$, or $2\gb \in I_\bbP$. 
Then  we will show how to relax this condition.

The second moment method is not applied directly to the partition function, but to 
  a {\sl reduced} version for which we allow at most one contact in $[0,1]$ and none in $(-\infty,0)$. 
Note in fact that
\begin{equation}
\label{eq:reducedZ}
\begin{split}
Z^{\gb,\go,\hat \phi}_{N,h,K}\, &\ge \, Z^{\gb,\go,\hat \phi}_{N,h,K}\left(\left\{ \phi:\, \sum_{x\in\tilde \gL_N} \delta_x \,\in \, \{0,1\} ,
\sum_{x\in\tilde \gL_N} \rho_x=0
\right\}\right)
\\
& = \,   
\bP_N^{\hat \phi} \left(\phi_x>1 \text{ for } x\in \tilde \gL_N \right)
\\
&\phantom{mo}+ \sum_{x\in\tilde \gL_N} e^{\gb \go_x -\gl(\gb) +h} 
\bP_N^{\hat \phi} \left(\gd_x =1 \text{ and } \sum_{y\in \tilde \gL_N\setminus \{x\}} \rho_y^+ =0\right)
\, =:\, Q^{\gb,\go,\hat \phi}_{N,h}
\,,
\end{split}
\end{equation}
and for conciseness we write $Q_N^{\go, \hat \phi}$ for $Q^{\gb,\go,\hat \phi}_{N,h}$: this is the reduced partition function.
Note that the reduced partition function does not contain $K$ and in fact \eqref{eq:reducedZ} holds uniformly in $K\ge 0$.

The first observation on $Q_N^{\go, \hat \phi}$ is that 
\begin{multline}
Q_N^{\go, \hat \phi} \ge   \bP_N^{\hat \phi} \left(\phi_x>1 \text{ for } x\in \tilde \gL_N \right) \\
=1-
 \bP_N^{\hat \phi} \left( \cup_{x\in \tilde \gL_N}\{ \phi_x \le 1\}\right) \ge  1- 
 \sum_{x\in \tilde \gL_N} \bP_N^{\hat \phi} \left(  \phi_x \le 1\right),
\end{multline}
and a direct estimate shows that, if $\hat \phi \in G_u$, we can find $c>0$ such that $$\bP_N^{\hat \phi} \left(  \phi_x \le 1\right)
\le \exp(-c (\log N)^2)$$ for every $x\in \tilde \gL_N$.  Hence, for $h$ sufficiently small we have 
$Q_N^{\go, \hat \phi} \ge 1/2$ and therefore
\begin{equation}
\label{eq:Taylor}
\log Q_N^{\go, \hat \phi}  \, \ge \, \left(Q_N^{\go, \hat \phi} -1\right) - \left(Q_N^{\go, \hat \phi} -1\right)^2\, .
\end{equation}
which leads to the bound (uniform in $K\ge 0$)
\begin{equation}
\label{eq:logLB}
\bbE \hat \bE^u \left[ \left(\log Z^{\gb,\go,\hat \phi}_{N,h,K}\right) \ind_{G_u}(\hat \phi)\right]\,
\ge \, \bbE \hat \bE^u \left[ \left(Q_N^{\go, \hat \phi} -1\right) \ind_{G_u}(\hat \phi)\right]-
\bbE \hat \bE^u \left[ \left(Q_N^{\go, \hat \phi} -1\right)^2 \ind_{G_u}(\hat \phi)\right]\,,
\end{equation}
on which we will concentrate our attention from here till the end of the proof.

\subsubsection{First moment estimates (lower and upper bound)}
Let us observe that $\bbE \left(Q_N^{\go, \hat \phi} -1\right)$ is equal to
\begin{equation}
\label{eq:st1fme}
 -\bP_N^{\hat \phi} \left( \cup_{x \in \tilde\gL_N} \{ \phi_x \le 1 \} \right) + e^h \sum_{x \in \tilde\gL_N}
\bP_N^{\hat \phi} \left( \gd_x=1 \text{ and } \phi_y>1 \text{ for } y \in  \tilde\gL_N\setminus \{x\}\right)\, , 
\end{equation}
and that this quantity, by the union bound, using also $e^h-1 \ge h$ and the notation $F_x:=\{\phi_y>1$  for  
$y \in  \tilde\gL_N\setminus \{x\}\}$, can be bounded below by
\begin{multline}
- \sum_{x \in \tilde\gL_N} \bP_N^{\hat \phi} \left(  \phi_x \le 1\right) +
\sum_{x \in \tilde\gL_N}
\bP_N^{\hat \phi} \left( \gd_x=1 \right)
- \sum_{x \in \tilde\gL_N}\bP_N^{\hat \phi} \left( \{\gd_x=1\}  \cap F_x^\complement  \right)
\\
+h  \sum_{x \in \tilde\gL_N} \bP_N^{\hat \phi} \left( \gd_x=1 \right)
-h  \sum_{x \in \tilde\gL_N} \bP_N^{\hat \phi} \left( \{\gd_x=1\} \cap F_x^\complement \right)\, ,
\end{multline}
which we reorder into
\begin{multline}
\label{eq:reorder}
\bbE \left(Q_N^{\go, \hat \phi} -1\right)\, \ge \, 
h  \sum_{x \in \tilde\gL_N} \bP_N^{\hat \phi} \left( \phi_x\le 1 \right)
- (1+h) \sum_{x \in \tilde\gL_N} \bP_N^{\hat \phi} \left( \phi_x<0 \right)
\\
- (1+h)
\sum_{x \in \tilde\gL_N} \bP_N^{\hat \phi} \left( \{\gd_x=1\} \cap F_x^\complement \right)\, .
\end{multline}
Now 
we observe that
\begin{equation}
 \bP_N^{\hat \phi} \left( \{\gd_x=1\} \cap F_x^\complement \right)\,\le\, 
 \bP_N^{\hat \phi} \left(\gd_x=1\right)
 \max_{z\in [0,1]}
 \sum_{y \in \tilde\gL_N\setminus \{x\}}\bP_N^{\hat \phi} \left( \phi_y \le 1 \,\vert\, \phi_x =z \right)\, ,
\end{equation}
and we use the fact that the probability that a random walk issued from $y$ hits $\partial \gL_N$ before 
visiting $x$ is larger than the probability that a random walk in $\bbZ^d$ issued from $y$  never 
hits $x$, and this latter probability $q$ is  positive. Hence the mean of $\phi_y$, under 
$\bP_N^{\hat \phi} \left( \cdot \,\vert\, \phi_x =z \right)$, is at least $qu/2$,
because 
$\hat \phi \in G_u$,
 and therefore, since the variance is 
bounded (by $\gs_d^2$),  there exists $c>0$ such that 
$\bP_N^{\hat \phi} \left( \phi_y \le 1 \,\vert\, \phi_x =z \right)\le \exp(-c (\log h)^2)$
and therefore the last term in \eqref{eq:reorder} is negligible with respect to the first in the right-hand side of the same formula.
Therefore for $\hat \phi \in G_u$ and for $h$ sufficiently small we have
\begin{equation}
\label{eq:reorder2}
\bbE   \left(Q_N^{\go, \hat \phi} -1\right)\, \ge \, 
\frac 45 h  \sum_{x \in \tilde\gL_N} \bP_N^{\hat \phi} \left( \phi_x\le 1 \right)
- (1+h) \sum_{x \in \tilde\gL_N} \bP_N^{\hat \phi} \left( \phi_x<0 \right)
\, .
\end{equation}
We will also need an upper bound on $\bbE   \left(Q_N^{\go, \hat \phi} -1\right)$.
For this we restart from
\eqref{eq:st1fme} and observe that
\begin{equation}
\label{eq:ubmean1}
\begin{split}
\bbE\left(Q_N^{\go, \hat \phi} -1\right)\, &= \, 
-\bP_N^{\hat \phi} \left( \cup_{x \in \tilde\gL_N} \{ \phi_x \le 1 \} \right) + e^h \sum_{x \in \tilde\gL_N}
\bP_N^{\hat \phi} \left( \left\{\gd_x=1\right\}\cap F_x\right)
\\
 & \le (e^h-1) \sum_{x \in \tilde\gL_N}
\bP_N^{\hat \phi} \left( \left\{\gd_x=1\right\}\cap F_x\right)
\\
& \le \,
(e^h-1) \sum_{x \in \tilde\gL_N}
\bP_N^{\hat \phi} \left( \gd_x=1\right) \,.
\end{split}
\end{equation}

\subsubsection{Second moment estimate}
Recall that we assume here that $\bbE [\xi^2]< \infty$. First of all
\begin{equation}
\bbE \left[\left(Q_N^{\go, \hat \phi} -1\right)^2 \right]\, =\, \left(\bbE\left(Q_N^{\go, \hat \phi} -1\right)\right)^2 
+ \textrm{var}_{\bbP}\left(Q_N^{\go, \hat \phi}\right).
\end{equation}
The variance term is easily computed and estimated:
\begin{equation}
\label{eq:variance}
\begin{split}
 \textrm{var}_{\bbP}\left(Q_N^{\go, \hat \phi}\right)
 \, &\le \, e^{2h} \Var(\xi)
 \sum_{x \in \tilde \gL_N} \bP_N^{\hat \phi} \left(\gd_x=1\right)^2\\
 &\le  e^{2h} \Var(\xi) \max_{x' \in \tilde \gL_N}  \bP_N^{\hat \phi} \left(\gd_{x'}=1\right)  \sum_{x \in \tilde \gL_N} \bP_N^{\hat \phi} \left(\gd_x=1\right)\,.
 \end{split}
 \end{equation}
For the square of the mean the estimate is already in \eqref{eq:ubmean1}. 
Hence
\begin{equation}\label{eq:variance2}
\begin{split}
\bbE \left[\left(Q_N^{\go, \hat \phi} -1\right)^2 \right]\, &\le \,
\left(e^{2h} \Var(\xi)+(e^h-1)^2 N^d\right) \max_{x' \in \tilde \gL_N}  \bP_N^{\hat \phi} \left(\gd_{x'}=1\right)  \sum_{x \in \tilde \gL_N} \bP_N^{\hat \phi} \left(\gd_x=1\right)
\\
& \le \, 2N^{d} \max_{x' \in \tilde \gL_N}  \bP_N^{\hat \phi} \left(\gd_{x'}=1\right)  \sum_{x \in \tilde \gL_N} \bP_N^{\hat \phi} \left(\gd_x=1\right)
\\
& \le \, \exp(-c (\log h)^2 )  \sum_{x \in \tilde \gL_N} \bP_N^{\hat \phi} \left(\gd_x=1\right)
\, .
\end{split}
\end{equation}
In both inequalities we used that  $h$ is  small (how small depends on $\Var (\xi)$: 
we require 
$e^{2h} \Var (\xi)\le N^d$ and $(e^h-1)^2 \le 1$) and that  $N=\exp(|\log h|^{3/2})$, and in the last inequality we used
$\hat \phi \in G_u$ (which ensures that the mean of $\phi_x$ under $\bP_N^{\hat \phi}$ is of order $|\log h|$). Nete that the constant $c$ does not depend on $\xi$. We have insisted on the role of $\xi$ to prepare the generalization to the case in which $\xi$ has unbounded second moment. 
\subsubsection{Lower bound on $\log Z$}
We go back to \eqref{eq:logLB}: uniformly in $K\ge 0$
\begin{multline}
\label{eq:logLB2}
\bbE \hat \bE^u \left[\left( \log Z^{\gb,\go,\hat \phi}_{N,h,K}\right) \ind_{G_u}\left(\hat \phi\right)\right]\,
\ge 
\,\frac 45 h  \sum_{x \in \tilde\gL_N} \hat \bE^u\left[\bP_N^{\hat \phi} \left( \phi_x\le 1 \right)  \ind_{G_u}\left(\hat \phi\right)\right]
\\ \phantom{move}
- (1+h) \big\vert \tilde\gL_N\big\vert \bP^u\left( \phi_0<0 \right)
-  \exp(-c (\log h)^2 ) \big\vert \tilde\gL_N\big\vert \bP^{u} \left(\gd_0=1\right)
\\
\ge \, 
\frac 45 h  N^d \bP^{u} \left( \phi_0\le 1 \right) 
-\frac 45 h  \sum_{x \in \tilde\gL_N} \hat \bE^u\left[\bP_N^{\hat \phi} \left( \phi_x\le 1 \right)
 \ind_{G_u^\complement}\left(\hat \phi\right)
 \right] \phantom{move}
\\
- (1+h)N^d \bP^u\left( \phi_0<0 \right) -  \exp(-c (\log h)^2 ) N^d \bP^{u} \left(\gd_0=1\right)
\\
\ge \, 
\frac 45 h N^d P(u) - \frac 45 h  \bE^u\left[ \sum_{x \in \tilde\gL_N} \rho_x^+  ; \, G_u^\complement\right] - h^{b} N^d P(u)
\,,
\end{multline}
with $b \in (1, \tilde a )$ (recall that $\tilde a >1$), and $h$ sufficiently small.
In the last inequality we have controlled from below the two terms in the line before the last one 
by $- h^{b} N^d P(u)$: this is because
 \eqref{eq:ratio} tells us $\bP^u\left( \phi_0<0 \right)=O(h^{\tilde a})\bP^u\left( \phi_0\le 1 \right)$, 
 so the first term in the line before the last one in \eqref{eq:logLB2} is much larger than the second and all this line is controlled
 as we  claimed.
The second term in the lats line of \eqref{eq:logLB2} has been already treated in \eqref{eq:fromUnB1} and we readily see that 
it is negligible with respect to the first.
Therefore we get to
\begin{equation}
\label{eq:logLBm}
\bbE \hat \bE^u \left[ \log Z^{\gb,\go,\hat \phi}_{N,h,K} ; \, G_u\right]\,
\ge \, \frac 23 h N^d P(u) \, ,
\end{equation}
for $h$ sufficiently small and by Lemma~\ref{th:Gc} for every $K \ge 0$
\begin{equation}
\label{eq:logLBm2}
\bbE \hat \bE^u \left[ \log Z^{\gb,\go,\hat \phi}_{N,h,K} \right]\,
\ge \, \left(\frac 23 h - C_d h^2(\gl(\gb) \vee K )\right) N^d P(u) \, ,
\end{equation}
so  the proof
of Proposition~\ref{th:lb-disorder}, assuming $\bbE [\xi ^2]< \infty$ is complete, for every $K>0$. For  the case $K=\infty$ 
we apply Lemma~\ref{eq:Klarge} -- recall also \eqref{eq:YY67} -- with $K=h^{-1/2}$
and  \eqref{eq:logLBm2}. 

\subsubsection{Relaxing  the assumption $E[\xi^2]<\infty$}
Let us assume now only that $\gb\in I_\bbP$, that is 
\eqref{gammatail}.We then  replace $\xi$ by $\xi_H:=\min(\xi,H)$ in the partition function. 
Of course we have $E[\xi_H]<1$. However if we rescale $u$ accordingly 
all the computations of the above remain valid if one chooses
\begin{equation}
h\,=\, -\log \bbE\left[\xi_H\right]+s \, ,
\end{equation}
with $s>0$ and  choose $N=\exp(|\log s|^{3/2})$. In this setup $s$  plays the role of $h$. We obtain in particular that 
there exists $s_0:=s_0(H,\gep)$ such that 
for $s<s_0$ 
\begin{equation}\label{eq:truncabound}
  \tf_K(\gb,-\log \bbE[\xi_H]+s)\,\ge\,   \exp  
 \left( -(1+\gep) \frac{\gs_d^2}2
  \left( \log \frac 1 s \right)^2
  \right)\, ,
\end{equation}
and from this we extract  that $h_c(\gb)\le -\log \bbE[\xi_H]$, which, sending  $H\to \infty$, yields $h_c(\gb)=0$.

\smallskip

To obtain a lower bound on the free energy we assume  that $\gb$ is in the interior of $I_\bbP$ and
we  make explicit the estimate by carefully  tracking the $H$ dependence in the lower bound proof.
Of course the first moment estimates do not depend on the value of $H$, and browsing the part involving the second moment, 
 we can check that the variance of $\xi$ only intervenes in \eqref{eq:variance}-\eqref{eq:variance2}.
 It suffices that
 \begin{equation}
 \exp(2h)  \Var(\xi_H)\, \le\,   N^{d}\, , 
 \end{equation}
and since trivially $ \Var(\xi_H)\le H^{2}$,  $N\ge H$  suffices. Recalling \eqref{eq:parameters},    if $s\le \exp \left( -(\log H)^{2/3} \right)$
 and  if $H$ is sufficiently large, how large depends on $\gep$,  \eqref{eq:truncabound} holds.
On the other hand we have from \eqref{gammatail}
\begin{equation}
 -\log \bbE(\xi_H)\, =\, -\log \left( 1-\int_H^\infty \bbP(\xi>t) \dd t \right)\, \le\, \frac C {\gamma -1} H^{-(\gamma-1)}\, .
\end{equation}
Hence for small $h$ we can fix $s=h/2$ and $H= \exp(|\log h|^{3/2})$ and  we deduce from \eqref{eq:truncabound}
that 
\begin{equation}
   \tf_K(\gb,h)\ge \tf_K(\gb,-\log \bbE[\xi_H]+h/2)\,\ge\,   \exp  
 \left( -(1+\gep) \frac{\gs_d^2}2
  \left( \log \frac 2 h \right)^2\right)\, ,
\end{equation}
and the proof of Proposition~\ref{th:lb-disorder} is now complete.
\qed

\medskip

\begin{rem}
\label{rem:xi}
A look at the previous argument shows that it goes through even weakening a little \eqref{eq:truncabound}, therefore including 
cases in which $\lim_{t\to \infty} t^\gamma \bbP(\xi>t) =\infty$, for any $\gamma >1$, but it is equal to zero for $\gamma=1$. This means that
in some cases it works also for $\gb$ at the boundary of $I_\bbP$. 
However if the tail decay is too weak 
(e.g. $\bbP(\xi\ge t)\ge (t (\log t))^{-1}(\log \log t)^{-2}$),  \eqref{eq:mainfe} does not hold as it can be seen by applying and adapting 
the upper bound argument present in \cite{cf:GL}.
\end{rem}

\section{Infinite volume limit: proof of Theorem~\ref{th:path}}
\label{sec:path}

We recall that for Theorem~\ref{th:path}, we have chosen $\gL_N=\{-N, \ldots , N\}^d$, and $\tilde \gL_N$ accordingly.
In this section we always assume that $\gb\in I_{\bbP}$. 

The first remark is that $\{\bP^{\gb,\go}_{N,h}\}_{N=1,2, \ldots}$ is increasing for the order induced by stochastic domination.
Later on we will use also the more general statement 
\begin{equation}\label{stochdom}
\bP^{\gb,\go}_{\gL,h} \le  \bP^{\gb,\go}_{\gL',h} \quad \text{ if } \gL \supset \gL'.
\end{equation}
where $\le$ stands here for stochastic domination.
Hence we can couple the family of random variables $\phi^N=\{\phi^N\}_{x \in \bbZ^d}$ with law $\bP^{\gb,\go}_{N,h}:=
\bP^{\gb,\go}_{N,h, \infty}$ in a way that $\phi^N$ increases with $N$.
Therefore for every local continuous function $f: [0, \infty]^{\bbZ^d}\to \bbR$ we have that for every $\go$
\begin{equation}
\label{wconv}
 \lim_{N\to \infty} \bE^{\gb,\go}_{N,h}\left[f(\phi)\right]\, =\,  \bE^{\gb,\go}_{h}\left[f(\phi)\right]\, ,
\end{equation}
and, by the Dominated Convergence Theorem, the same holds by taking the $\bbE$ expectation on both sides
(we are using $\bE^{\gb,\go}_{h}$ for $\bE^{\gb,\go}_{\infty,h}$).

We are now going to argue  that $\bP^{\gb,\go}_{h}$ satisfies the Markov property. We use the notation $\cF_A$
for the $\gs$-algebra generated by $\phi_A:=\{\phi_x\}_{x \in A}$, $A\subset \bbZ^d$.
The aim is showing that for every finite subset $\Gamma$ of $\bbZ^d$   and for every local bounded continuous $g:
[0, \infty]^{\Gamma}\to \bbR$ -- in particular, the limit of $g(\phi_\gG)$, when $\phi_x\to \infty$ for every $x\in \gG$, exists and we call it $g(\infty)$ --  for all $\go\in \bbR^{\bbZ^d}$ we have that $\bP^{\gb,\go}_{h}(\dd \tilde\phi)$-a.s.
\begin{multline}
\label{eq:markovinfvol}
  \bE^{\gb,\go}_{h}\left[ g\left( \phi_\Gamma\right)\, \big\vert \,
  \cF_{\gG^\complement}
  \right]\left(\tilde \phi\right)\,=\\ \begin{cases}
  \frac{1}{Z^{\gb,\go, \tilde\phi, +}_{\gG,h}}  \bE^{\tilde \phi,+}_{\gG} 
  \left[\exp\left(\sum_{x\in \gG} (\gb \go-\gl(\gb)+h) \delta_x\right) g\left( \phi_\Gamma\right)
    \right] & \text{ if } \max \phi_{\partial^+\Gamma} < \infty\, ,\\
    g(\infty) & \text{ if } \max \phi_{\partial^+\Gamma}  = \infty \,,
    \end{cases}
    \end{multline}
where $\partial^+ \Gamma:= \{y\in \Gamma^\complement:$ there exists $x \in \Gamma$ such that $y \sim x \}$ and 
$ \bP^{\tilde\phi,+}_{\gG}$ is the law of a free field $ \phi$ with boundary condition $\tilde \phi$ on $\bbZ^d \setminus \gG$,
recall \eqref{density}, conditioned to $\{\phi:\, \phi_x \ge 0$ for every $x \in \gG\}$. 
$Z^{\gb,\go,\tilde\phi,+}_{\gG,h}$ is the obvious normalization constant associated to the Boltzmann term that appears in the right-hand side.

Call $G_{g, \Gamma}(\tilde\phi)$ the right-hand side of \eqref{eq:markovinfvol}. Two important observations are:
\smallskip

\begin{enumerate}
\item $G_{g, \Gamma}(\tilde\phi)$ depends only on $\phi_{\partial^+ \gG}$: this is  the Markov property. We will then 
consider $G_{g, \Gamma}(\cdot)$ as a function from $[0, \infty]^{\partial^+ \gG}$ to $\bbR$. 
\item Of course $ \Vert G_{g, \Gamma}\Vert_\infty \le \Vert g \Vert_\infty$ and one directly verifies also the continuity 
of $G_{g, \Gamma}(\cdot)$.
\end{enumerate}
\smallskip

To prove \eqref{eq:markovinfvol} it suffices to show that for every bounded local continuous $f: [0, \infty]^{\Gamma^\complement}\to \bbR$
we have that 
\begin{equation}
\label{eq:markovinfvol1}
 \bE^{\gb,\go}_{h}\left[ f\left( \phi_{\Gamma^\complement}\right)g\left( \phi_\Gamma\right)\right]\, =\, 
 \bE^{\gb,\go}_{h}\left[ f\left( \phi_{\Gamma^\complement}\right)G_{g, \Gamma}(\phi)
 \right]\,.
\end{equation}
But, by continuity and boundedness of the integrands, in both sides of \eqref{eq:markovinfvol1} we can replace 
$\bE^{\gb,\go}_{h}[\ldots]$ with $\lim_N\bE^{N, \gb,\go}_{h}[\ldots]$ and the finite volume statement is directly verified
as soon as $N$ is sufficiently large, since the locality of $f$ implies that $f(\phi)=f(\phi')$ if $\phi_{\gL_N}=\phi'_{\gL_N}$
for $N$  larger than a finite value that depends on $f$. So \eqref{eq:markovinfvol} holds and the infinite volume field we built 
satisfies the Markov property.

\medskip

Next we prove that the quenched measure $\bP^{\gb,\go}_h$ is translationally covariant and two results about the quenched averaged measure $\bbE \bP^{\gb,\go}_h$. Translation covariance for the quenched limit probability and 
translation invariance of  $\bbE \bP^{\gb,\go}_h$ stem from the same argument, that we give now. Using
 \eqref{stochdom} one checks that that for $x\in \bbZ^d$, we have the following stochastic comparison for the translated measure for finite $N> \vert x\vert$:
\begin{equation}
\label{eq:fortransl}
 \bP^{\gb,\go}_{N-|x|,h} \, \le\,  \bP^{\gb,\go}_{N,h}\Theta_x\, \le\,   \bP^{\gb,\go}_{N+|x|,h}\, ,
\end{equation}
where $ \bE^{\gb,\go}_{N,h}\Theta_x[h(\phi)]= \bE^{\gb,\go}_{N,h}[h(\Theta_x\phi)]$ for every bounded local continuous function $h$. Translation covariance of the quenched measure follows by taking $N \to \infty$,  translation invariance for  the quenched averaged measure follows by taking the $\bbE$ expectation of the three terms in \eqref{eq:fortransl}
and by sending $N \to \infty$. 

For the second result on the quenched averaged measure let $\partial_h^-$ and $\partial_h^+$ denote, respectively,  the left and right derivative with respect to $h$.

\medskip

\begin{lemma}
\label{th:contact}
For every $h$  
\begin{equation}
\label{eq:contact}
\bbE \bE^{\gb,\go}_{h}\left[ \gd_0 \right] \, \in \, \left[\partial_h^- \tf_\infty(\gb, h), \partial_h^+ \tf_\infty(\gb, h)\right]\, .
\end{equation}
\end{lemma}
\medskip

\begin{lemma}
\label{th:ergodic}
For every $h$, if $\{\phi_x\}_{x \in \bbZ^d}$ is distributed according to  $\bbE \bP^{\gb,\go}_h$,
the random field $\{\gd_x\}_{x \in \bbZ^d}$,   is (translation) ergodic.
\end{lemma}

\medskip

Let us see how these two lemmas, and the Markov property,  allow to conclude the proof. 
First of all ergodicity implies that 
\begin{equation}
\label{eq:0-1}
\bbE \bP^{\gb,\go}_{h}\left( 
\text{there exists } x \in \bbZ^d:\, \gd_x=1 \right) 
\, \in \, \{0,1\}\, .
\end{equation}
Of course  $\bbE \bE^{\gb,\go}_{h}\left[ \gd_0 \right]$ is either zero or positive: Lemma~\ref{th:contact}
ensures that this dichotomy precisely corresponds to the localization transition, that is to $h \le 0$ and $h>0$ by Theorem~\ref{th:mainfe}.
It also corresponds to the dichotomy \eqref{eq:0-1}
by elementary arguments. 

Consider first the case  $h>0$, that is  in the case in which the probability in \eqref{eq:0-1} is equal to one, 
and therefore 
\begin{equation}
\label{eq:forabs0}
\bP^{\gb,\go}_{h}\left( 
\text{there exists } x \in \bbZ^d:\, \gd_x=1 \right)\, =\, 1\, , \ \ \ \ \bbP(\dd \go)-\text{a.s.}\, .
\end{equation}
We claim that, $\bbP(\dd \go)$-a.s., $\bP^{\gb,\go}_{h}( \phi_y=\infty)=0$ for every $y$, hence that
$\bP^{\gb,\go}_{h}( $there exists $y$ such that $\phi_y=\infty)=0$. In fact, reasoning by absurd, 
 if   there exists $y$ such that 
 $\bP^{\gb,\go}_{h}( \phi_y=\infty)>0$ then, by the Markov property \eqref{eq:markovinfvol},   for every $x\neq y$ we have 
 $\bP^{\gb,\go}_{h}( \phi_x=\infty, \phi_y=\infty)= \bP^{\gb,\go}_{h}( \phi_y=\infty)>0$, so, iterating countably many times the argument,  
 we see that, $\bbP(\dd \go)$-a.s.,  $\bP^{\gb,\go}_{h}( \phi_y=\infty$ for every $ y \in \bbZ^d)>0$,
 which contradicts the statement \eqref{eq:forabs0}. Therefore the claim is proven and therefore we have also
 that $\bbE \bP^{\gb,\go}_{h}($there exists $y$ such that $\phi_y=\infty)=0$. 
 
 On the other hand if $h \le 0$ we are in the case in which the probability in \eqref{eq:0-1}
is equal to zero. Hence 
\begin{equation}
\label{eq:forabs1}
\bP^{\gb,\go}_{h}\left( 
\text{there exists } x \in \bbZ^d:\, \gd_x=1 \right)\, =\, 0\ \ \ \ \bbP(\dd \go)-\text{a.s.}
\end{equation}
In particular for the same $\go$'s for  any $x$ we have that 
$\bP^{\gb,\go}_{h}( \gd_x=1)=0$. By the Markov property \eqref{eq:markovinfvol} this implies that $\phi_y=\infty$ for at least a $y \sim x$.
But we have just seen from the previous argument that  $\bP^{\gb,\go}_{h}(\phi_x=\infty$ for every $x)=
\bP^{\gb,\go}_{h}(\phi_y=\infty)$, which in this case is one. 

The proof of Theorem~\ref{th:path} is complete.
\qed

\medskip

\noindent
{\it Proof of Lemma~\ref{th:contact}}
Set $m=\bbE \bE^{\gb,\go}_{h}[\gd_0]$. We recall that $\bbE \bE^{\gb, \go}_{h, N} [\gd_0]$ decreases as $N$ grows.
The limit is $m$ by convergence in law,  cf. \eqref{wconv}, because the discontinuity point of $\gd_0$ is $\phi_0=1$ and 
$\bP^{\gb, \go}_{h, N}(\phi_0=1)=0$ as one can see by conditioning on  $\cF_{\{0\}^\complement}$ and using 
\eqref{eq:markovinfvol}: if the $\phi$ values on which we condition on the nearest neighbors of $0$ are all finite than 
the conditional measure has a density, otherwise the field at the origin takes the value $\infty$. Either ways this conditional probability is zero and the claim follows.
By exploiting further the monotonicity under set inclusion of the measure we directly  
 see that 
that for every $\gep>0$ we can find $N_0$ such that for $N > N_0$
\begin{equation}
\bbE \bE^{\gb, \go}_{h, N} [\gd_x] \, \in \, [m, m+\gep]\, ,
\end{equation}
for every $x\in \tilde \gL_{N-N_0}$. But then 
\begin{equation}
\partial _h \bbE\log Z_{N, h, \infty}^{\gb, \go} \, =\, 
  \bbE \bE^{\gb,\go}_{h,N}\left [\sum_{x\in \tilde \gL_N} \delta_x\right]\, \le\,  \left|\tilde \gL_{N-N_0}\right| (m+\gep)+ \left|\tilde \gL_N \setminus \tilde \gL_{N-N_0}\right|\, ,
\end{equation}
and therefore the superior limit as $N \to \infty$ of the left-hand side, normalized by $\vert \tilde \gL_N\vert$,  it is not larger than $m+\gep$.
Similarly 
\begin{equation}
\partial _h \bbE\log Z_{N, h, \infty}^{\gb, \go} \, =\, 
  \bbE \bE^{\gb,\go}_{h,N}\left [\sum_{x\in \tilde \gL_N} \delta_x\right]\, \ge \,  \left|\tilde \gL_{N-N_0}\right| m\, ,
\end{equation}
and  the inferior limit of this left hand-side, normalized by $\vert \tilde \gL_N\vert$, is not smaller than $m$. 
\qed 

\medskip

\noindent
{\it Proof of Lemma~\ref{th:ergodic}}
Let $A$ be a translation invariant event in the $\gs$-algebra generated by 
$\{\gd_x\}_{x \in \bbZ^d}$. We can approximate the event by $A_M$ which just depends on $\{\gd_x\}_{x\in \tilde \gL_M}$ 
in a way that
\begin{equation}
\label{simdif}
 \bbE \bP^{\gb,\go}_{h} \left(A\Delta A_M\right)\,\le\,  \gep\, .
\end{equation}
Furthermore we can choose $N>M$ so large  that 
\begin{equation}
\label{inequin}
\bbE \bE^{\gb,\go}_{N,h}\left[\sum_{x\in \tilde \gL_M} \delta_x\right] - \bbE \bE^{\gb,\go}_{h}\left[\sum_{x\in \tilde \gL_M} \delta_x\right]\, \le\,  \gep\, .
\end{equation}
This is a consequence of the convergence of the sequence of measures, cf. \eqref{wconv},  because 
$\bbE \bP^{\gb,\go}_{h}(\cup_{x \in \tilde \gL_M} \{\phi_x=1\})=0$ 
as can be seen  
by a conditioning argument like in the very beginning of the proof of Lemma \ref{th:contact}.

 Now set $v_N$ be a vector with all entries $0$ except one that is equal to $3N$.
\begin{equation}
\gG(N)\, :=\,  \gL_{N}\cup \Theta_{v_N} \gL_N.
\end{equation}
Note that $\gG_N$ is composed of two disjoint boxes and thus that under 
$\bP^{\gb,\go}_{\gG(N),h}$, $\{\phi_x\}_{x\in \gL_N}$ and $\{\phi_x\}_{x\in \Theta_{v_N}\gL_N}$ are independent, so 
\begin{multline}\label{eq:factorize}
\bbE\bP^{\gb,\go}_{\gG(N),h}\left[A_M \cap\Theta_{v_N} A_M \right]
= \bbE\left[ \bP^{\gb,\go}_{\gL_N,h}(A_M) \bP^{\gb,\go}_{\theta_{v_N}\gL_N,h}(\Theta_{v_N} A_M) \right]\\
= \bbE\left[ \bP^{\gb,\go}_{\gL_N,h}[A_M]\right] \bbE \left[ \bP^{\gb,\go}_{\theta_{v_N}\gL_N,h}(\Theta_{v_N} A_M) \right]
= \left(\bbE\left[ \bP^{\gb,\go}_{\gL_N,h}[A_M]\right]\right)^2.
\end{multline}
where in the first equality we used the Markov property and for the second  we used independence of the environment in the two boxes.

We assume  $N>2M$ so that  $\tilde \gL_M\cup \Theta_{v_N}\tilde \gL_M\subset \gG(N)$ and the distance of 
both $\gL_M$ and $\Theta_{v_N}\tilde \gL_M$ to the boundary of $\gG(N)$ is more than $N/2$. 
We have
\begin{equation}
\label{inequi}
 \bbE \bE^{\gb,\go}_{\gG(N),h} \left[ \sum_{x\in \tilde \gL_M\cup \Theta_{v_N}\tilde \gL_M} \delta_x \right] - 
 \bbE \bE^{\gb,\go}_{h} \left[\sum_{x\in \tilde \gL_M\cup \Theta_{v_N}\tilde \gL_M} \delta_x
 \right]\,\le\, 2\gep\, ,
\end{equation}
as can be directly extracted froom \eqref{inequin} because the first addendum in the left-hand side can be written as 
the sum of two terms on on which we can apply \eqref{inequin} after using translation invariance. 
Now by stochastic domination we know that there exists a monotone coupling between the two probabilities.
For such a coupling $\{\delta^1_{x}\}_{x\in\tilde \gL_M\cup \Theta_{v_N}\tilde \gL_M}$ and $\{\delta^2_{x}\}_{x\in\tilde \gL_M\cup \Theta_{v_N}\tilde \gL_M}$ 
coincide with probability at least $2\gep$ (we use Markov's inequality together with \eqref{inequi}).
As a consequence we have 
\begin{equation}
\label{kdo}
  \left|\bbE \bP^{\gb,\go}_{\gG(N),h}
  \left(A_M \cap \Theta_{v_N}A_M\right)
  -  \bbE \bP^{\gb,\go}_{h}\left(A_M \cap \Theta_{v_N}A_M\right) \right|\, \le\, 2\gep\, .
\end{equation}
Note that in the same manner as for \eqref{inequin} -- the boundary of $A_M$ is  a subset of
$\cup_{x \in \tilde \gL_M} \{\phi_x=1\}$ --  we have also that for $M$ sufficiently large 
\begin{equation}\label{kdod}
 \left|\bbE \bP^{\gb,\go}_{N,h}(A_M)-  \bbE\bP^{\gb,\go}_{h}(A_M) \right|\, \le\,  \gep\, .
\end{equation}
By putting  everything together (using the triangle inequality and \eqref{eq:factorize}) we obtain
\begin{equation}
\begin{split}
\Big|
\bbE \bP^{\gb,\go}(A)&-\bbE \bP^{\gb,\go}(A)^2\Big|
=
\left|\bbE \bP^{\gb,\go}_{h}(A \cap \Theta_{v_N}A)- \bP^{\gb,\go}(A)^2\right|\\
\le &\left|\bbE \bP^{\gb,\go}_{h}(A \cap \Theta_{v_N}A)-\bbE \bP^{\gb,\go}_{h}(A_M \cap \Theta_{v_N}A_M)\right|\\
&+\left|\bbE \bP^{\gb,\go}_{h}(A_M \cap \Theta_{v_N}A_M)-\bbE \bP^{\gb,\go}_{\gG(N),h}(A_M \cap \Theta_{v_N}A_M)\right|\\
& +\left|\bbE \bP^{\gb,\go}_{N,h}(A_M)^2- \bbE \bP^{\gb,\go}_{h}(A_M)^2\right|
 +\left|\bbE \bP^{\gb,\go}_{h}(A_M)^2-\bbE \bP^{\gb,\go}_{h}(A)^2\right|\le 8\gep.
\end{split}
\end{equation}
The last inequality comes from the fact that all four terms are smaller than $2\gep$, the first from \eqref{simdif} and translation invariance, the second  from \eqref{kdo}, the third from \eqref{kdod}
and the last one from \eqref{simdif}.
Since $\gep>0$ is arbitrary we obtain that $\bbE \bP^{\gb,\go}(A)\in\{0,1\}$.
\qed

\medskip

{\bf Acknowledgements:} H.~L. acknowledges the support of a productivity grant from CNPq.

 \appendix
 
 \section{Free energy: existence and other estimates}
 \label{sec:fe}
 
 \begin{theorem}
 \label{th:fe}
 For every $K\in (-\infty, \infty]$, every $\gb \in I_\bbP$ and every $h\in \bbR$ we have
 that the  limit
 \begin{equation}
 \label{eq:fe}
  \lim_{N \to \infty} \frac 1{N^d}  \log Z^{\gb,\go,0}_{N,h,K}\, .
 \end{equation}
 exists $\bbP(\dd \go)$-a.s. and in $L^1(\bbP)$ and the limit is not random.
 \end{theorem}
 \medskip
 
 Therefore \eqref{eq:fe-qa} provides a definition of $\tf_K(\gb, h)$.
 
 \medskip
 
 \noindent
 {\it Proof.}
 As long as $K$ is finite the arguments in \cite{cf:CM1} go through and they yield the result. For 
 $K=\infty$ we observe that, since the partition function decreases as $K$ increases for every $K$
 (so, in particular, $F_\infty(\gb, h)= \lim_{K \to \infty} F_K(\gb, h)\ge 0$)
 \begin{equation}
 \limsup_{N \to \infty}
 \frac 1{N^d }\log Z_{N, h, \infty}^{\gb,\go,0}\, \le \, \tf_K(\gb, h)\, ,
 \end{equation}
 $\bbP(\dd \go)$-a.s.. On the other hand Lemma~\ref{th:Klarge} ensures  that
 \begin{equation}
 \liminf_{N \to \infty}
 \frac 1{N^d }\log Z_{N, h, \infty}^{\gb,\go,0}\, \ge \, \tf_K(\gb, h)- r(K)\, ,
 \end{equation}
 with $\lim_{K \to \infty}r(K)=0$ and this gives \eqref{eq:fe}
 in the  $\bbP(\dd \go)$-a.s. sense. The $L^1(\bbP)$ limit can then be obtained by an application
 of the Dominated Convergence Theorem.
 \qed

 \medskip
 
 As an important technical tool 
 we have the following analog of \cite[Prop.~4.2]{cf:GL}: the proof is a direct generalization because the 
 potential terms are bounded.
 
 \medskip
 
 \begin{proposition}
 \label{th:superadd}
 For any value of $u\in \bbR$, $K\in \bbR$, $h\in \bbR$ and $\gb \in I_\bbP$ we have 
 \begin{equation}
 \label{eq:limit} 
 \lim_{N\to \infty} \frac{1}{N^d} \bbE\hat \bE^u\left[ \log Z^{\gb,\go,\hat \phi}_{N,h,K}  \right]\,  = \, \tf_K(\gb,h)\, .
 \end{equation}
 Moreover for any $u$ and $N$ one has 
 \begin{equation}
 \label{eq:superadd}
 \frac{1}{N^d} \bbE\hat \bE^u\left[ \log Z^{\gb,\go,\hat \phi}_{N,h,K}  \right] \, \le \, \tf_K(\gb,h)\, .
 \end{equation}
 \end{proposition}
 \medskip

 \medskip
 
 \begin{lemma}
 \label{th:Klarge}
 For every $K$,  $h$ and $\gb \in I_\bbP$ we have that the bound
 \begin{equation}
 \label{eq:Klarge}
 \liminf_{N \to \infty}
 \frac 1{N^d }\log Z_{N, h, \infty}^{\gb,\go,0}
  \, \ge \, \tf_K(\gb, h) - \bbE\left[\log\left(1+\exp\left(-K+(\gb\go_1-\gl(\gb)+h)_-\right)\right)\right]\, ,
 \end{equation}
 holds $\bbP(\dd \go)$-a.s.. Moreover \eqref{eq:Klarge} still holds 
 if $\log Z_{N, h, \infty}^{\gb,\go,0}$ in the left-hand side is replaced by  $\bbE \log Z_{N, h, \infty}^{\gb,\go,0}$. 
 \end{lemma}
 \medskip
 
  Of course $\bbE\left[\log\left(1+\exp\left(-K+(\gb\go_1-\gl(\gb)+h)_-\right)\right)\right]=o(1)$ as $K \to \infty$ 
  by the Dominated Convergence Theorem, but the estimate is quantitative. 
 In fact for every $\gb\ge 0$  and $h \ge 0$ we can find $c=c_\gb>0$ such that for every $K$ sufficiently large we have
 \begin{equation}
 \label{eq:YY67}
 \bbE\left[\log\left(1+\exp\left(-K+(\gb\go_1-\gl(\gb)+h)_-\right)\right)\right]\, \le \, \exp(-c_\gb K)\,.
 \end{equation}
 In fact, it is immediate to see that $c_0=1$. For $\gb>0$ it is sufficient to argue for $h=0$ and with $Y=\gb\go_1-\gl(\gb)$ we have
  \begin{multline}
 \label{eq:Y67}
 \bbE\left[\log\left(1+\exp\left(-K+(Y)_-\right)\right)\right]
 \, \le\\
 \log\left(1+\exp\left(-K/2)\right)\right)
 +
  \bbE\left[\log\left(1+\exp\left(-K+(Y)_-\right)\right); Y< -K/2\right]
  \\
  \le \, \log\left(1+\exp\left(-K/2\right)\right)
 +
  \bbE\left[
  \left(
  \log 2+ (Y)_-\right); Y< -K/2\right]\,,
 \end{multline}
 and observe that, since by assumption there exists $a>0$ such that $\gl(-a \gb)< \infty$ 
 \begin{equation}
 \bbP\left(Y< -K/2\right)\, =\, \bbP\left(-a\gb \go_1 > \frac a2 K- a\gl(\gb) \right)\, \le 
 \, \exp\left( \gl(-a\gb)+a \gl(\gb)-\frac a2K\right)\, .
 \end{equation} 
 The conclusion, that is \eqref{eq:YY67}, is now obtained by applying the Cauchy-Schwarz inequality 
 to the very last term in \eqref{eq:Y67}.
 
 \medskip
 
 \noindent
 {\it Proof.}
 We start with observing that the left-hand side in \eqref{eq:Klarge} is $\bbP(\dd\go)$-a.s. equal to
 \begin{equation}
  \tf_K(\gb,h) + \liminf_{N \to \infty} \frac 1 {N^d}\log \bP^{\gb,\go,0}_{N, h,K}\left(\phi_x\ge 0 \text{ for every } x \in \mathring{\gL}_N\right) \, ,
 \end{equation}
 and we have to bound from below the inferior limit in the last expression.
 For this we observe that if we set $E_A^-:=\{\phi\in \bbR^{\mathring{\gL}_N}: \, \phi_x<0$ for $x\in A$ and $\phi_x\ge 0$ for every $x\in \mathring{\gL}_N \setminus A\}$, we have (with the concise notation $Y_x:=\gb \go_x-\gl(\gb) +h$)
 \begin{equation}
  \bP^{\gb,\go,0}_{N, h,K}\left(E_A^-\right)\, =\, \exp(-K \vert A \vert) \int_{E_A^-} \frac{\exp\left( \sum_{x\in \tilde \gL_N}
  Y_x \gd_x\right)}
  {Z_{N,h,K}^{\gb,\go,0}} \bP_N^0(\dd \phi)\, ,
 \end{equation}
 and by performing the change of variables $\tilde\phi_x = -\phi_x$ if $x \in A$ and $\tilde\phi_x=\phi_x$ otherwise, we see that
 \begin{equation}
  \int_{E_A^-}{\exp\left( \sum_{x\in \tilde \gL_N} Y_x\gd_x\right)}
   \bP_N^0(\dd \phi)\, \le \, \exp\left(\sum_{x\in A} Y_x\right)\int_{E_\emptyset ^-}{\exp\left(\sum_{x\in \tilde \gL_N}Y_x \gd_x\right)}
   \bP_N^0(\dd \phi)\, ,
 \end{equation}
 because such a transformation has (absolute value) of the Jacobian determinant equal to one, 
 $\sum_{x,y} (\tilde \phi_x - \tilde \phi_y)^2 \le \sum_{x,y} (\phi_x - \phi_y)^2$, where the sums are over
 the nearest neighbor $(x,y)$ in $\gL_N^2\setminus (\partial \gL_N)^2$, and 
 $\sum_{x\in \tilde \gL_N} Y_x \gd_x$   under such transformation can decrease of at most $\sum_{x \in A} (Y_x)_-$. Since of course 
 $E_\emptyset^-=\{\phi: \, \phi_x\ge 0$ for every $x\in \mathring{\gL}_N\}$ and since 
 $\sum_{A\subset \mathring{\gL}_N}  \bP^{\gb,\go,0}_{N, h,K}\left(E_A^-\right)=1$ we see that
 \begin{equation}
 1\, \le \, \left(\sum_{A\subset \mathring{\gL}_N}\prod_{x \in A}\exp\left(-K+(Y_x)_- \right) \right)  \bP^{\gb,\go,0}_{N, h,K}\left(\phi_x\ge 0 \text{ for every } x  \in\mathring{\gL}_N \right)\, ,
 \end{equation}
 and since the sum is equal to $\prod_{x \in \mathring{\gL}_N}(1+\exp(-K+(Y_x)_-))$, the claim in Lemma~\ref{th:Klarge} follows
 by applying the law of large numbers to the family of $L^1$ IID random variables 
 $\left\{ \log (1+\exp(-K+(Y_x)_-)) \right\}_{x\in \bbZ^d}$.
 \qed


\begin{thebibliography}{99}


 \bibitem{cf:AZ}
 K.~S.~Alexander and N.~Zygouras, \emph{Equality of critical points for polymer depinning transitions with loop exponent one}, Ann. Appl. Probab. {\bf 20} (2010), 356-366.   
   
   



\bibitem{cf:BDZ}      E.~Bolthausen, J.-D.~Deuschel and O.~Zeitouni, \emph{Entropic repulsion of the lattice free field},   Commun. Math. Phys. {\bf 170} (1995),  417-443.
 
     
\bibitem{cf:BDZwetting}      E.~Bolthausen, J.-D.~Deuschel and O.~Zeitouni, \emph{Absence of a wetting transition for a pinned harmonic crystal in dimensions three and larger},   J. Math. Phys. {\bf 41} (2000),  1211-1223.



\bibitem{cf:CV}
P. Caputo and Y. Velenik, 
\emph{A note on wetting transition for gradient field},  Stoch. Proc. Appl. {\bf 87} (2000), 107-113.


 \bibitem{cf:CCH}
 A.~Chiarini, A.~Cipriani and R.~S.~Hazra, \emph{A note on the extremal process of the supercritical Gaussian Free Field},
		Electron. Comm. Probab. {\bf 20} (2015) 74.
%
%
     
\bibitem{cf:CM1} L. Coquille and
    P. Milos,  \emph{A note on the discrete Gaussian free field with disordered pinning on $\bbZ^d$, $d\ge 2$},
Stoch.  Proc. and  Appl. {\bf 123} (2013) 3542-3559.
    

\bibitem{cf:D}
J.-D. Deuschel, \emph{Entropic repulsion of the lattice free field. II. The $0$-boundary case},  Commun. Math. Phys. {\bf 181} (1996),  647-665.


\bibitem{cf:DG}
J.-D. Deuschel and G. Giacomin,
\emph{Entropic repulsion for the free field: pathwise behavior in $d\ge 3$},
Comm. Math. Phys. {\bf 206} (1999), 447-462.  
  
   
 \bibitem{cf:GB} G. Giacomin, {\sl Random polymer models}, 
Imperial College Press, World Scientific (2007). 

 
\bibitem{cf:G} G. Giacomin, \emph{Disorder and critical phenomena through basic probability models}, \'Ecole d'\'et\'e de probablit\'es de Saint-Flour XL-2010, Lecture Notes in Mathematics {\bf 2025}, Springer, 2011.

\bibitem{cf:GL} G.~Giacomin and H.~Lacoin,
\emph{Pinning and disorder relevance for the lattice Gaussian free field},
arXiv:1501.07909, to appear on JEMS

 
 \bibitem{cf:Hcrit} A.~B.~Harris, {\it Effect of random defects on the critical behaviour of Ising models} J. Phys. C {\bf 7} (1974),
1671-1692.

%
  
   \bibitem{cf:FF2}
 H.~Lacoin, \emph{Pinning and disorder for the Gaussian free field II: the two dimensional case},  arXiv:1512.05240 [math-ph].


\bibitem{cf:LM} J.~L.~Lebowitz and C.~Maes, \emph{The effect of an external field on an interface, entropic repulsion}, J. Statist. Phys. {\bf 46} (1987),  39-49. 

\bibitem{cf:sohier} J.~Sohier,
\emph{The scaling limits of the non critical strip wetting model}, 
Stoch. Proc. Appl. {\bf 125}  (2015), 3075-3103.
  
 \bibitem{cf:Vel} Y.~Velenik, 
\emph{Localization and delocalization of random interfaces},  
Probab. Surv. {\bf 3} (2006), 112-169. 

\bibitem{cf:ofer}
O.~Zeitouni, \emph{Branching random walks and Gaussian fields}, lecture notes,  available on the webpage of the author
   
 
\end{thebibliography}
\end{document}